\documentstyle[12pt,aasms4,graphics]{article}
\newcommand{\etal}{et~al.}

\begin{document}
\oddsidemargin=0mm

\title{
The Unique Type Ia Supernova 2000cx in NGC 524
}

\author{Weidong Li\altaffilmark{1}, Alexei V. Filippenko\altaffilmark{1},
Elinor Gates\altaffilmark{2}, Ryan Chornock\altaffilmark{1},
Avishay Gal-Yam\altaffilmark{3,4}, Eran O. Ofek\altaffilmark{3}, 
Douglas C. Leonard\altaffilmark{5},
Maryam Modjaz\altaffilmark{1},
R. Michael Rich\altaffilmark{6},
Adam G. Riess\altaffilmark{7},
 and Richard R. Treffers\altaffilmark{1} \\
Email: wli@astro.berkeley.edu, alex@astro.berkeley.edu,
egates@ucolick.org}

\altaffiltext{1}{Department of Astronomy, University of California, Berkeley, CA 94720-3411.}

\altaffiltext{2}{Lick Observatory, PO Box 82, Mount Hamilton, CA 95140.}

\altaffiltext{3}{School of Physics and Astronomy, and the Wise Observatory, Tel Aviv University, Israel.}

\altaffiltext{4}{Colton Fellow.}

\altaffiltext{5}{Five College Astronomy Department, University of Massachusetts,
Amherst, MA 01003-9305.}

\altaffiltext{6}{Department of Physics and Astronomy, University of California,
Los Angeles, CA 90095-1562.}

\altaffiltext{7}{Space Telescope Science Institute, 3700 San Martin Drive, Baltimore, MD 21218.}

\begin{abstract}

We present extensive photometric and spectroscopic observations of the Type Ia
supernova (SN Ia) 2000cx in the S0 galaxy NGC 524, which reveal it to be
peculiar.  Photometrically, SN 2000cx is different from all known SNe Ia, and
its light curves cannot be fit well by the fitting techniques currently
available. There is an apparent asymmetry in the $B$-band peak, in which the
premaximum brightening is relatively fast (similar to that of the normal SN
1994D), but the postmaximum decline is relatively slow (similar to that of the
overluminous SN 1991T).  The color evolution of SN 2000cx is also peculiar: the
$(B - V)_0$ color has a unique plateau phase and the $(V - R)_0$ and $(V -
I)_0$ colors are very blue.

  Although the premaximum spectra of SN 2000cx are similar to those of SN
1991T-like objects (with weak Si~II lines), its overall spectral evolution is
quite different.  The Si~II lines that emerged near maximum $B$-band brightness
stay strong in SN 2000cx until about three weeks past maximum. The change in
the excitation stages of iron-peak elements is slow. Both the iron-peak and the
intermediate-mass elements are found to be moving at very high expansion
velocities in the ejecta of SN 2000cx.

  We discuss theoretical models for SN 2000cx. SN 2000cx may be an overluminous
object like SN 1991T, but with a larger yield of $^{56}$Ni and a higher kinetic
energy in the ejecta. We also briefly discuss the implications of our
observations for the luminosity vs. light-curve width relation.

\end{abstract}


\section{INTRODUCTION}

Studies of high-redshift SNe Ia have revealed a surprising cosmological result,
that the expansion of the Universe is presently accelerating due to a nonzero
cosmological constant (e.g., Riess et al. 1998a, 2001; Perlmutter et
al. 1999). This result, however, is based on the assumption that there are no
significant differences between SNe Ia at high redshift and their low-redshift
counterparts. Although this assumption is supported to first order by
comparisons of the photometric and spectroscopic properties of SNe~Ia at high
and low redshifts (e.g., Riess et al. 1998a, 2000; Coil et al. 2000), there is
also some evidence for differences between SNe Ia at different redshifts (e.g.,
Riess et al. 1999; Falco et al. 1999; Aldering, Knop, \& Nugent 2000; Howell,
Wang, \& Wheeler 2000; Li et al. 2001). It is thus important to verify this
assumption by further observations of high-redshift and nearby SNe Ia.

High-quality observations of nearby SNe Ia provide valuable information
about their progenitor evolution and the relevant physics.  Analyses of 
samples of well-observed nearby SN Ia enable observers to study the 
differences among SNe Ia, empirical correlations,  and possible environmental
effects (e.g., Hamuy et al. 2000).  It is thus important to expand the sample
of well-observed nearby SNe Ia. 

SN 2000cx in the S0 galaxy NGC 524 was discovered and confirmed by the Lick
Observatory Supernova Search (LOSS; Treffers et al. 1997; Li et al. 2000;
Filippenko et al. 2001) with the 0.76~m Katzman Automatic Imaging Telescope
(KAIT) in unfiltered images taken on 2000 July 17.5 and 18.4 UT (Yu, Modjaz, \&
Li 2000; UT dates are used throughout this paper). A low-resolution optical
spectrum obtained with the Lick Observatory 1-m Nickel telescope on July 23
(Chornock et al. 2000) identified the event as a peculiar SN 1991T-like type
Ia, with prominent Fe III absorption troughs near 4300 \AA\, and 4900 \AA\, but
weak Si II at 6150 \AA.  Yu, Modjaz, \& Li (2000) measured an accurate position
for SN 2000cx as $\alpha$ = 1$^h$24$^m$46$^s$.15, $\delta = +9\arcdeg 30\arcmin
30\farcs 9$ (equinox J2000.0), which is $23\farcs 0$ west and $109\farcs3$
south of the nucleus of NGC 524. The SN was at magnitude 14.5 (unfiltered) at
discovery, and subsequently became the brightest SN of the year 2000.

Shortly after the discovery of SN 2000cx, a follow-up program of multicolor
photometry and spectroscopy was established at Lick Observatory.  Photometry of
SN 2000cx was also gathered at the Wise Observatory (WO) in Israel.  This paper
presents the results from these campaigns and is organized as follows. Section
2 contains a description of the observations and analysis of the photometry,
including our methods of performing photometry, our calibration of the
measurements onto the standard Johnson-Cousins system, our resulting multicolor
light curves, and our comparisons between the light curves and color curves of
SN 2000cx and those of other SNe Ia. Section 3 contains a description of the
spectral observations and analysis. The good temporal coverage of the spectral
observations enables us to monitor the evolution of many features. We also
undertake a thorough comparison between the spectra of SN 2000cx and those of
other SNe Ia. We discuss the implications of our observations in \S 4 and
summarize our conclusions in \S 5.

\section{PHOTOMETRY}

\subsection{Observations and Data Reduction}

Broadband $UBVRI$ images of SN 2000cx were obtained using an Apogee AP7 CCD
camera with KAIT, but only the $BVRI$ observations are reported in this
paper. (The $U$-band data will be combined with observations of other SNe Ia
and reported in a future paper.) The Apogee camera has a back-illuminated SITe
512$\times$512 pixel CCD chip with UV2AR coating to boost the blue response. At
the $f/8.2$ Cassegrain focus of KAIT, the 24 $\mu m$ pixel of the chip yields a
scale of 0$\farcs$8 pixel$^{-1}$, making the total field of view of the camera
6$\arcmin.7\times 6\arcmin.7$. The typical seeing at KAIT is around 3$\arcsec$
full width at half maximum (FWHM), so the CCD images are well sampled.

The WO images of SN 2000cx were obtained with the Wise 1-m telescope in the
$f/7$ configuration, as part of the WO queue-observing program. The CCD camera
with the back-illuminated 1024$\times$1024 pixel Tektronix chip was used in
unbinned mode. With a pixel scale of about 0$\arcsec.$7 pixel$^{-1}$, the total
field of view was $\sim$ 12$\arcmin\times12\arcmin$.  The typical images have
FWHM $\approx$ 2$\arcsec.$5 and thus are also well sampled.

The bias and dark-current subtraction with subsequent twilight-sky flatfielding
were accomplished automatically at each of the telescopes.  Significant
fringing in the $I$-band KAIT images could not be completely removed, probably
because the near-infrared sky lines vary in brightness with time. No detectable
fringing is seen in the WO images. The implication of the fringing for the
photometry will be discussed in more detail later (\S 2.2).

Figure 1 shows a KAIT $V$-band image taken on 2000 July 25 with SN 2000cx and
four local standard stars marked. Absolute calibration of the field was done
with KAIT on 2000 September 5, 25, and 26 by observing Landolt (1992) standard
stars at different airmasses throughout the photometric nights. Instrumental
magnitudes for the standard stars were measured using aperture photometry with
the IRAF\footnote{IRAF (Image Reduction and Analysis Facility) is distributed
by the National Optical Astronomy Observatories, which are operated by the
Association of Universities for Research in Astronomy, Inc., under cooperative
agreement with the National Science Foundation.}  DAOPHOT package (Stetson
1987) and then used to determine transformation coefficients to the standard
system of Johnson et al. (1966, for $BV$) and Cousins (1981, for $RI$). The
derived transformation coefficients and color terms were then used to calibrate
the sequence of four local standard stars in the SN 2000cx field. The
magnitudes of those four stars and the associated uncertainties derived by
averaging over the three photometric nights are listed in Table 1. The field of
SN 2000cx was not calibrated at WO, so the WO images were also reduced against
these local standard stars.

\renewcommand{\arraystretch}{0.45}

\begin{deluxetable}{lllll}
\tablecaption{Photometry of comparison stars}
\tablehead{
\colhead{ID}&\colhead{$V$}&
\colhead{$(B-V)$}&\colhead{$(V-R)$}&
\colhead{$(V-I)$} 
}
\startdata
1 & 12.555(007) & 0.586(009) & 0.378(006) & 0.760(007) \\
2 & 13.506(008) & 0.797(009) & 0.477(006) & 0.921(008) \\
3 & 12.800(007) & 0.696(008) & 0.406(007) & 0.794(007) \\
4 & 14.540(018) & 0.829(008) & 0.459(010) & 0.887(009) \\
\enddata    
\tablenotetext{}{Note: uncertainties are indicated in parentheses.}
\end{deluxetable}

We used the point-spread-function (PSF) fitting method (Stetson 1987)
to perform differential photometry of SN 2000cx relative to the comparison stars. 
Although SN 2000cx is at a fortunate position (far from the nucleus
of the host galaxy,  with a faint and smoothly varying background) and 
simple aperture photometry might suffice, we chose to use PSF fitting
for the following two reasons. (1) PSF fitting usually gives more 
accurate measurements than aperture photometry for objects projected 
upon slightly more complex regions. It also has the advantage of 
creating an image with all the measured objects subtracted, 
which can serve as an indicator of the quality of the magnitude
measurements of the comparison stars and the SN. (2) PSF fitting 
handles the fringing in the $I$-band KAIT images better than 
aperture photometry does, especially at late times when the SN was 
rather faint. 

Before the photometry measurements were initiated, cosmic rays were carefully
removed from all the images, and dark-current residuals were manually
eliminated from the KAIT images. This latter step is taken because the KAIT CCD
camera is cooled thermoelectrically rather than with liquid nitrogen, so the
temperature of the camera is not very stable and dark current is not always
cleanly subtracted from the images. The resulting low-level dark-current
residual varies from image to image and must be removed manually by adding or
subtracting a few percent of the long-exposure dark-current image. The
uncertainties introduced by the manual scaling and removing of the dark current
have only negligible effects on the magnitude measurements of the high-contrast
SN and the local standard stars.  The WO CCD camera is cooled by liquid
nitrogen, and there is no dark-current residual problem in the images.

The PSF for each image was determined with DAOPHOT using Stars 1, 3, and 4
(Star 2 is a bit too close to the host galaxy), which are bright and
well-isolated. To increase the signal-to-noise ratio of the differential
photometry, we used only the inner core (with a radius comparable to the FWHM)
of SN 2000cx and the local standard stars to fit the PSF. The sky backgrounds
of the stars were measured using an annulus with an inner radius of 20 pixels
(16$\arcsec$ in the KAIT images and 14$\arcsec$ in the WO images) and an outer
radius of 25 pixels (20$\arcsec$ in the KAIT images and 17$\arcsec$.5 in the WO
images).

The standard Johnson-Cousins $BVRI$ magnitudes of SN 2000cx
were obtained by doing differential photometry between the 
instrumental magnitudes of the SN and those of the four local
standard stars. The color terms involved in this transformation
are $-$0.06, 0.04, 0.08, and $-$0.01 for the KAIT $B, V, R,$ and 
$I$ filters,  respectively. Notice that these color terms are quite
different from those used by Modjaz et al. (2000) for the KAIT data
on SN 1998de, especially in the $R$ band. This is caused by our 
installation of a new set of filters for KAIT in 1999, whose 
transmissivities were made to match those defined  by Bessell (1990)
for the Johnson-Cousins $UBVRI$ photometric system. The color
terms for the WO 1-m $B, V, R, $ and $I$ filters are $-$0.11, $-$0.02,
$-$0.02, and $-$0.04,  respectively.

\begin{deluxetable}{ccccc}
\tablecaption{KAIT photometry of SN 2000cx}
\tablehead{
\colhead{JD $-$} & \colhead{$B$} & \colhead{ $V$ } &
\colhead{$R$} & \colhead{$I$}  \\
\colhead{2450000} & \colhead{(mag)} &
\colhead{(mag)} &
\colhead{(mag)} &
\colhead{(mag)} 
}
\startdata
1743.97&14.41(03)&14.23(04)&14.17(03)&14.25(03) \\
1744.96&14.17(02)&14.00(02)&13.97(03)&14.10(03) \\
1746.00&13.95(02)&13.80(02)&13.78(02)&13.88(02) \\
1746.99&13.79(02)&13.65(02)&13.68(04)&13.78(04) \\
1749.95&13.48(05)&13.33(08)&13.35(09)&$-$ \\
1750.95&13.45(02)&13.34(02)&13.39(03)&13.63(02) \\
1752.96&13.44(02)&13.27(02)&13.36(02)&13.72(02) \\
1753.96&13.45(02)&13.27(02)&13.36(02)&13.77(02) \\
1754.96&13.49(02)&13.25(02)&13.37(02)&13.81(02) \\
1756.92&13.58(02)&13.30(02)&13.40(02)&13.88(02) \\
1757.96&13.65(02)&13.31(02)&13.47(03)&14.03(04) \\
1758.96&13.73(02)&13.38(02)&13.55(02)&14.08(02) \\
1759.96&13.77(03)&13.42(02)&13.66(03)&14.13(03) \\
1760.99&13.84(03)&13.46(02)&13.69(02)&14.27(03) \\
1761.99&13.97(02)&13.56(02)&13.82(03)&14.44(05) \\
1762.99&14.02(02)&13.62(02)&13.94(02)&14.52(04) \\
1764.94&14.16(02)&13.78(02)&14.14(03)&14.64(06) \\
1766.94&14.32(02)&13.91(02)&14.24(04)&14.72(05) \\
1768.98&14.54(03)&14.07(02)&14.32(04)&14.72(03) \\
1770.92&14.72(03)&14.16(02)&14.36(03)&14.66(03) \\
1772.92&14.94(02)&14.27(03)&14.35(04)&14.58(03) \\
1778.93&15.64(03)&14.77(02)&14.54(03)&14.49(04) \\
1780.89&15.85(03)&14.93(02)&14.70(04)&14.57(06) \\
1783.92&16.10(03)&15.22(03)&14.90(04)&14.76(03) \\
1789.96&16.43(06)&15.51(05)&15.24(06)&15.28(13) \\
1791.88&16.48(02)&15.59(03)&15.33(02)&15.36(02) \\
1792.80&16.50(03)&15.62(03)&15.37(03)&15.38(03) \\
1794.90&16.56(03)&15.72(03)&15.47(03)&15.54(03) \\
1796.86&16.59(03)&15.78(02)&15.52(04)&15.64(03) \\
1805.92&16.79(04)&16.10(03)&15.89(03)&16.17(05) \\
1810.95&16.93(03)&16.27(02)&16.09(04)&$-$ \\
1812.86&16.92(03)&16.33(03)&16.19(05)&16.48(05) \\
1813.80&16.95(03)&16.33(03)&16.15(03)&16.47(04) \\
1817.90&17.03(04)&16.48(03)&16.32(03)&16.75(04) \\
1821.88&17.08(03)&16.63(03)&16.47(04)&16.94(05) \\
1825.84&17.18(04)&16.71(03)&16.59(03)&17.09(05) \\
1832.87&17.35(12)&$-$&$-$&$-$ \\
1836.82&17.37(04)&16.99(03)&16.96(03)&17.48(10) \\
1840.87&17.49(05)&17.14(03)&17.12(06)&17.84(16) \\
1845.88&17.60(10)&17.29(09)&17.36(09)&$-$ \\
1849.87&17.85(28)&17.31(13)&17.46(08)&18.32(24) \\
1853.82&17.73(06)&17.50(04)&17.49(04)&18.31(09) \\
1861.82&17.85(08)&17.67(08)&17.71(09)&18.54(15) \\
1865.79&18.00(09)&17.74(07)&17.84(06)&18.49(10) \\
1873.70&18.06(05)&17.89(05)&18.03(05)&18.77(11)\\
1881.67&18.23(05)&18.03(05)&18.18(06)&19.15(16)\\
1889.65&18.36(10)&18.17(11)&$-$      &19.06(17)\\
1897.67&18.59(08)&$-$      &18.80(14)&$-$      \\
1905.58&18.73(08)&18.54(07)&19.00(13)&$-$      \\
\enddata    
\tablenotetext{}{Note: uncertainties are indicated in parentheses.}
\end{deluxetable}

\newpage

\begin{deluxetable}{ccccc}
\tablecaption{Wise Observatory photometry of SN 2000cx}
\tablehead{
\colhead{JD $-$} & \colhead{$B$} & \colhead{ $V$ } &
\colhead{$R$} & \colhead{$I$}  \\
\colhead{2450000} & \colhead{(mag)} &
\colhead{(mag)} &
\colhead{(mag)} &
\colhead{(mag)} 
}
\startdata
1753.53&13.42(02)&13.25(02)&13.34(03)&13.76(02)\\
1754.56&13.44(03)&13.23(02)&13.32(02)&13.78(02)\\
1755.48&13.49(04)&13.23(03)&13.35(03)&13.81(04)\\
1760.44&13.81(02)&13.43(02)&13.62(03)&14.20(03)\\
1762.51&13.94(02)&13.59(02)&13.85(03)&14.39(02)\\
1767.47&14.36(02)&13.96(03)&14.20(03)&14.60(03)\\
1774.46&15.11(05)&14.40(04)&14.35(04)&14.44(05)\\
1777.53&$-$&14.62(05)&14.47(03)&14.41(04)\\
1778.56&15.61(03)&14.75(05)&14.51(03)&14.39(03)\\
1794.33&16.53(05)&15.67(03)&15.45(04)&15.42(03)\\
1795.48&16.62(03)&15.75(02)&15.51(03)&15.52(03)\\
1801.38&16.76(08)&15.99(05)&15.75(05)&15.89(05)\\
1809.44&16.87(03)&16.23(03)&16.03(03)&16.22(03)\\
1810.45&16.88(03)&16.22(03)&16.05(04)&16.31(04)\\
1811.34&16.93(03)&16.25(03)&16.09(03)&16.42(05)\\
1819.27&17.12(06)&16.53(04)&16.39(03)&16.71(05)\\
1821.42&17.12(03)&16.60(03)&16.45(03)&$-$\\
1822.33&17.15(04)&16.60(03)&16.49(03)&16.92(05)\\
1837.32&17.42(05)&17.06(04)&17.00(05)&17.42(12)\\
1842.24&17.47(05)&17.17(04)&17.25(05)&17.78(25)\\
1845.27&17.53(04)&17.30(03)&17.26(03)&17.63(05)\\
1846.27&17.54(04)&17.26(02)&17.26(05)&17.89(04)\\
1856.29&$-$&17.52(04)&17.68(05)&18.04(09)\\
1864.28&$-$&17.71(03)&17.78(04)&18.37(06)\\
1865.32&$-$&17.75(03)&17.82(03)&18.44(05)\\
1870.30&$-$&17.87(03)&17.98(03)&18.50(05)\\
1878.33&$-$&18.02(03)&18.18(03)&18.86(10)\\
\enddata    
\tablenotetext{}{Note: uncertainties are indicated in parentheses.}
\end{deluxetable}

Our final $BVRI$ measurements of SN 2000cx
are listed in Table 2 for KAIT and Table 3 for the WO telescope.
Uncertainties for the measurements were estimated by combining in quadrature
the errors given by the photometry routines in DAOPHOT with those 
introduced by the transformation of instrumental magnitudes onto
the standard system. 

\subsection{Optical Light Curves}

\subsubsection{Overall Results}

Figure 2 displays our $BVRI$ light curves of SN 2000cx. The KAIT data points
are shown with open circles and the WO ones with solid circles. For most of the
points the uncertainties are smaller than the plotted symbols. The overall
agreement between the measurements from the two telescopes is excellent. When
KAIT data points are linearly interpolated to estimate photometry at the epochs
of the WO observations, the difference between the KAIT and the WO photometry
is usually less than 0.04 mag.  However, there seem to be some differences in
the $I$ band, in particular between JD 2,451,765 and 2,451,780 and after JD
2,451,840. The SN was quite bright in the $I$ band ($<$14.8 mag) between JD
2,451,765 and 2,451,780, so the offset between the measurements, with WO being
brighter by about 0.1 mag, may be caused by the differences in the CCD infrared
response and the $I$-band filter transmission curve of the two systems (i.e.,
even after the transformation onto the standard system using proper color
terms, as discussed below).  The SN became fainter than 17.5 mag in the $I$
band after JD 2,451,840, and the fringing in the $I$-band KAIT images makes it
difficult to obtain accurate magnitudes even with PSF fitting. Compared with
the WO measurements, the KAIT data show larger fluctuations in the light
curve. For this reason, we consider the WO measurements for the late-time
$I$-band photometry to be more representative of SN 2000cx than the KAIT ones.

Kevin Krisciunas (2001, private communication) made available to us the
photometry of SN 2000cx obtained at the Apache Point Observatory (APO)
and the Manastash Ridge Observatory (MRO). A comparison between
the KAIT and the APO/MRO photometry of SN 2000cx indicates that they
agree with each other to 0.02 mag except in the $I$ band, 
where the APO/MRO measurements are systematically brighter
by about 0.05 mag.

Suntzeff (2000) found that there was a systematic
difference of about 0.07 mag in the $BV$ passbands between two
data sets of SN 1998bu obtained with two telescopes
at Cerro Tololo Inter-American Observatory (CTIO), and cautioned the
need for photometrists to correct their measurements  to a standard
filter transmission curve using spectrophotometry.  We found that for the
case of SN 2000cx, the $BVR$ measurements from different sources 
agree with each other quite well,  probably due to similar filter
transmission and CCD quantum efficiency in those passbands at different
telescopes.
Different CCD chips
usually have somewhat different infrared responses, and the $I$-band
filter transmission curves are quite diverse, 
so it is not surprising
that there are systematic differences among the $I$-band data sets
of SN 2000cx. In any case, it seems that if a telescope has 
non-standard Johnson-Cousins filters, or if the CCD camera used
for photometry has poor quantum efficiencies in certain passbands,
it may be necessary to follow the suggestion by Suntzeff (2000) 
to correct the photometry using spectrophotometry, in a manner
similar to $K$-corrections (e.g., Jha et al. 1999). 

Here, and in all subsequent discussions and figures, the KAIT and
the WO datasets of SN 2000cx are combined and discussed together.
Also defined here is the variable $t$, which is the time since 
maximum brightness in the $B$ band (JD = 2451752.2; see discussions
below).

\begin{deluxetable}{ccccc}
\tablecaption{Photometric information on SN 2000cx}
\tablehead{
\colhead{Filter} &\colhead{$B$} & \colhead{$V$} &
\colhead{$R$} & \colhead{$I$} }
\startdata

UT of max. &Jul 26.7 $\pm$ 0.1&Jul 28.8 $\pm$ 0.1&Jul 28.1 $\pm$ 0.1&Jul 25.1 $\pm$ 0.5 \\
Julian Date of max.&2451752.2 $\pm$ 0.1&2451754.3 $\pm$ 0.1&245173.6 $\pm$ 0.1 &2451750.6 $\pm$ 0.5 \\
Magnitude at max.& 13.43 $\pm$ 0.03&13.25 $\pm$ 0.03&13.37 $\pm$ 0.04&13.62 $\pm$ 0.04\\
$\Delta m_{15}$ &0.93 $\pm$ 0.04 &0.82 $\pm$ 0.05&0.94 $\pm$ 0.05 &1.06 $\pm$ 0.06 \\
\enddata
\end{deluxetable}

The date and the magnitude of the peak in each passband
are listed in Table 4; they were determined by fitting different 
spline functions and polynomials to the $BVRI$ light curves around maximum
brightness. Since we have nearly nightly observations of SN 2000cx from
$t$ = $-$8 to 20 days,  the measurements from
different fitting methods are quite consistent, yielding comparatively
small uncertainties for them. Also listed in Table 4 are the 
$\Delta m_{15}$ values for all the passbands. In accordance with 
Phillips (1993), $\Delta m_{15}(X)$ is 
defined as the decline (in magnitudes) during the first 15
days after maximum brightness in the passband $X$. As pointed out by 
Phillips (1993) and Hamuy et al. (1996a), these $\Delta m_{15}$ values,
especially $\Delta m_{15}(B)$, are good indicators of the luminosities
of SNe Ia in the sense that a SN Ia with a smaller $\Delta m_{15}$ value
(also called a ``slow decliner") is more luminous than a faster decliner.
This light-curve shape and luminosity correlation is often referred as the
luminosity -- $\Delta m_{15}(B)$ or the luminosity vs. light-curve
width relation for SNe Ia.

In Figures 3, 4, 5, and 6, we compare the light curves of SN 2000cx
with those of several other well-observed SNe Ia representing the 
diversity of SN Ia light curves: SN 1991T [$\Delta m_{15}(B)=0.95\pm0.05$, 
Lira et al. 1998] as an example of an overluminous SN Ia; SN 1992bc
[$\Delta m_{15}(B)=0.87\pm0.05$, Hamuy et al. 1996b], SN 1992al [$\Delta
m_{15}(B)=1.11\pm0.05$, Hamuy et al. 1996b], and SN 1994D [$\Delta m_{15}(B)
=1.31\pm0.08$, Richmond et al. 1995] as examples of normal SNe Ia with
different decline speeds in their light curves; and SN 1991bg
[$\Delta m_{15}(B)=1.93\pm0.08$, Filippenko et al. 1992a;
Leibundgut et al. 1993] as an example of a subluminous SN Ia. 
All light curves are shifted in time and peak magnitude to match
those of SN 2000cx with the time zero-point being the date of maximum
light in the $B$ band. 

\subsubsection{$B$ Band}

The $B$-band light curve of SN 2000cx (Figure 3) shows a peculiar 
evolution, unlike that of any other SN in the comparison. Before maximum
brightness, SN 2000cx brightens much faster than SN 1991T, although 
its spectra show strong similarity to those of SN 1991T
at this time (see \S 3 for more details of spectral comparisons).
In fact, the rise of SN 2000cx is just like that of SN 1994D, whose spectra
show apparent differences from those of SN 2000cx. Thus, the premaximum
brightening of SN 2000cx provides 
a counterexample to the correlation between spectroscopic peculiarity
and light-curve behavior of SNe Ia (Nugent et al. 1995;
Riess et al. 1998b).

The light curve of SN 2000cx follows that of SN 1994D until $t$ = 6 days, 
but then it has a  transition phase where the decline slows
down (more easily seen in the inset of Figure 3) 
and changes to follow the light curve of SN 1991T from $t$ = 15 to 22
days. Because of this transition phase, SN 2000cx
has a $\Delta m_{15}(B)$ value of $0.93 \pm 0.04$ mag, very similar to
that of SN 1991T ($0.95 \pm 0.05$ mag). There is an apparent asymmetry
in the rising and declining parts of the light curve for
SN 2000cx, which might challenge the light-curve
fitting techniques currently available for SNe Ia. For example, 
both the ``stretch method" (Perlmutter et al. 1997) and 
the multicolor light-curve shape method (hereafter MLCS; Riess,
Press, \& Kirshner 1996) require all 
SNe Ia to constitute one family of well-behaved light curves. A similar,
but less apparent asymmetry can be seen in the light curve 
of SN 1992bc (Figure 3). These asymmetries indicate that $\Delta m_{15}(B)$ 
may not be the  only parameter characterizing the width of 
light curves of SNe Ia, and raise the question of whether SN Ia luminosities
are correlated more with their $\Delta m_{15}(B)$ values or 
their light-curve widths.

There seems to be another transition between $t$ = 22 and 30
days, during which time SN 2000cx declines faster (0.11 mag day$^{-1}$) than 
SN 1991T (0.09 mag day$^{-1}$), SN 1992bc (0.09 mag day$^{-1}$), and SN 1992al (0.10 mag day$^{-1}$).
By $t$ = 30 days, SN 2000cx begins to follow SN 1992al
and continues until $t$ = 60 days.  At late times all SNe Ia in 
Figure 3 show a linear decline, with the decline rate measured in
mag (100 days)$^{-1}$ sorted from the slowest to the fastest as follows:
SN 1991T, 1.41; SN 1992bc, 1.49; SN 1992al, 1.54; SN 1994D, 1.77; 
SN 2000cx, 1.92; and SN 1991bg, 2.35. SN 2000cx has the second
fastest decline rate. 

\subsubsection{$V$ Band}

The $V$-band light curve of SN 2000cx is shown in Figure 4. SN 2000cx
behaves like SN 1994D until $t$ = 4 days, and then evolves in a manner between
SN 1992al and SN 1994D until $t$ = 30 days. The excellent temporal 
coverage and high accuracy of the SN 2000cx photometry reveal 
a ``shoulder" phase
at $t$ = 16 to 30 days.  Considering the peculiar photometric
evolution of SN 2000cx, however, this shoulder phase may
be unique to SN 2000cx and not common for all SNe Ia. 
After $t$ = 30 days 
the light curve of SN 2000cx falls below that of SN 1994D, and joins
that of SN 1991bg. It is remarkable that despite the huge difference
in the light-curve shapes 
between SN 2000cx and SN 1991bg 
around peak brightness, the two SNe have
a very similar total decline in the first 40 days after peak (2.45 mag for 
SN 2000cx and 2.47 mag for SN 1991bg). 

Between $t$ = 50 and 100 days all SNe show a linear decline, with the 
decline rates in mag (100 days)$^{-1}$ measured as follows: SN 1991T, 2.29;
SN 1992bc, 2.47; SN 1992al, 2.63; SN 1994D, 2.74; SN 2000cx, 2.82;
and SN 1991bg, 3.33. SN 2000cx again shows a faster decline than most
of the SNe, second only to SN 1991bg.

The late-time $V$-band light curve of SN 2000cx ($t >$ 100 days) is well
observed. It seems that $t$ = 110 days is a bending
point for SN 2000cx,  where the decline rate changes from 2.82 mag (100 days)$^{-1}$
to 1.95 mag (100 days)$^{-1}$. This bending point, however, is much less dramatic
than the one at $t \approx$ 35 days,  and an alternative view is that
the decline rate changes gradually over time.
A change in the late-time decline rate
is also found for SN 1991T (Lira et al. 1998) and SN 1991bg (Turatto et al. 
1996), so it may be common in SNe Ia. 

There is a general consensus that the late-time light curves of SNe Ia are
powered by the radioactive decay energy of $^{56}$Co to $^{56}$Fe. 
Milne, The, \& Leising (2001) used Monte Carlo algorithms to simulate
the interactions of the products from this decay, $\gamma$-ray photons and
positrons, and confirmed that the late-time
light curves of SNe Ia can be explained with energy deposition
from those $\gamma$-ray photons and positrons { only if substantial
positrons escape from the magnetic field of the ejecta}.
They also suggested that at late times ($t >$ 40 days),
the $V$-band flux constitutes a constant fraction ($\sim$25\%) of, and thus
scales as,
the bolometric luminosity of a SN Ia. For SN 2000cx,  the 
latest (and smallest) decline rate of 1.95 mag (100 days)$^{-1}$ in the 
$V$ band (and thus in bolometric luminosity) is still steeper 
than that expected from the $^{56}$Co to $^{56}$Fe decay with 
instantaneous deposition of all decay energy [0.98 mag (100 days)$^{-1}$],
suggesting that the $\gamma$-ray photon and positron transport are
indeed important at those late times.

\subsubsection{$R$ Band}

The $R$-band light curve comparison is presented in Figure 5. SN 2000cx shows
an evolution similar to that of SN 1994D, especially before 
$t$ = 0 day and at $t$ = 30 -- 70 days. However, there are apparent
differences between the two SNe from $t$ = 0 to 30 days. As can be seen
more clearly in the inset of Figure 5, SN 2000cx has 
a slower decline rate, but a larger overall decline after peak than SN 1994D
does. The plateau phase begins at $t$ = 12 days for SN 1994D, while it 
begins at $t$ = 16 days for SN 2000cx. Unfortunately, there are no 
data available for SN 1994D after $t$ = 70 days, so it is unclear
whether the late-time evolution of SN 2000cx and SN 1994D is similar. 
The late-time photometry of SN 2000cx indicates that there may be 
a bending point at $t \approx$ 100 days, but the changes in the 
decline rate may be gradual. 

\subsubsection{$I$ Band}

Figure 6 shows the $I$-band light curves of the SNe. SN 2000cx behaves
like SN 1994D before $t$ = 7 days, but then it declines faster and more
than SN 1994D. The minimum of the ``dip" between the two peaks of
the $I$-band
light curves occurs at about 0.6 mag below peak at $t \approx$ 10 days for 
SN 1994D, while it occurs at about 1.1 mag below peak at $t \approx$ 14 days
for SN 2000cx. 
SN 2000cx has the most prominent ``dip" among all the SNe
Ia in the comparison. Pinto \& Eastman (2000) examined
the nature of the opacity and radiation transport in SNe Ia, and 
explored the reasons for those secondary maxima in the $I$ band.
They found that several weeks after maximum brightness, 
the ionization in regions of trapped radiation falls to 
include significant amounts of singly ionized species (Ca II, Fe II, and Co II)
that can emit strongly in the near infrared. As a result, 
there is a decrease in the flux
mean opacity, which reduces the diffusion time,  allows the 
reservoir of trapped radiation to escape more rapidly,  and leads
to an increase in the luminosity. The more prominent ``dip" and fainter 
secondary peak of SN 2000cx may thus indicate
a slower evolution in the ionization stages of the species, which
is consistent with the spectral evolution of SN 2000cx discussed
in \S 3. 

All SNe in Figure 6 show a linear decline after
their second peak, with the decline rates in mag (100 days)$^{-1}$ measured
as follows: SN 1992bc, 3.51; SN 1991bg, 3.66; SN 1992al, 3.78;
SN 1991T, 4.05; SN 1994D, 4.71; and SN 2000cx, 4.79.  SN 2000cx thus has
the fastest decline rate after the second peak among all the SNe Ia
in the comparison. The decline rate of SN 2000cx 
seems to change at $t \approx$ 100
days, though the large scatter in the late-time photometry 
prevents us from making a definitive conclusion. 

In summary, the light curves of SN 2000cx differ from those of all
known SNe Ia. The closest match seems to be those of SN 1994D, especially
in the $R$ and $I$ bands, but there remain significant differences
between them. The ``metamorphosis" in the $B$ band (i.e., SN 2000cx
looks like SNe 1994D, 1991T, and 1992al at different epochs) is puzzling, 
as are the large postmaximum decline in the $V$ band and the secondary
maximum evolution in the $R$ and $I$ bands. 

\subsection{Optical Color Curves}

Strong host-galaxy reddening is not expected for SN 2000cx, since NGC 524 
is an early-type galaxy of Hubble type S0. Moreover, SN 2000cx is located
about 2$^\prime$ southwest of the nucleus (Figure 1) where no significant 
contamination is expected. Another argument against the presence of substantial
host-galaxy reddening is that the spectra of SN 2000cx do not show signs 
of narrow interstellar Na I D absorption at the expected position (5940 \AA). 
In most of the spectra, however, there is an apparent absorption feature
at about
5890 \AA\, (with equivalent width 1.1 $\pm$ 0.4 \AA) caused by
interstellar Na I within the Milky Way, indicating the existence
of Galactic reddening toward SN 2000cx. Using the full-sky maps of dust
infrared emission of Schlegel, Finkbeiner, \& Davis (1998), we derive
a Galactic reddening of $E(B - V) = 0.08$ mag for SN 2000cx. In subsequent
discussions, we assume no host-galaxy extinction and use $E(B - V) = 0.08$ mag as
the adopted total reddening for SN 2000cx. 

In Figures 7, 8, and 9, we present the intrinsic optical color curves of 
SN 2000cx [$(B - V)_0$, $(V - R)_0$, and $(V - I)_0$], together with color
curves of several other SNe Ia (1991T, 1992bc, 1992al, and 1994D) for
comparison. In order to calculate the color excess for the $R$ and $I$
passbands we adopt the standard reddening-law parameterization,
$E(V - R)/E(B - V) = 0.78$ and $E(V - I)/E(B - V) = 1.60$
(Savage \& Mathis 1979). 
The reddenings adopted for the comparison SNe are $E(B - V)$ = 0.13 mag
for SN 1991T (Phillips et al. 1992), 0.02 mag for SN 1992bc (Phillips
et al. 1999), 0.04 mag for SN 1992al (Phillips et al. 1999), and 0.04 mag
for SN 1994D (Richmond et al. 1995). 

The color evolution of SN 2000cx is peculiar. The $(B - V)_0$ color curve
in Figure 7 shows that SN 2000cx has $(B - V)_0$ = 0.10 mag at $t$ = $-$8 days,
the reddest of all the SNe in the comparison sample. In fact, SN 2000cx
has the reddest color among all the SNe in Figure 7 until $t \approx$ 10 days.
SN 2000cx has a unique phase of nearly constant color at 
$(B - V)_0 \approx 0.3$ mag at $t$ = 6 -- 15 days, which corresponds
to the transition phase seen in the $B$-band light curve (Figure 3).
Because of this unusual plateau phase, by $t$ = 15 days SN 2000cx has 
the second {\it bluest} color (exceeded only by SN 1992bc) among all
the SNe in Figure 7. All these SNe  show an approximately
linear increase in color between $t$ = 15 and 30 days, though the slopes
differ somewhat (SN 2000cx has the smallest one). 

Lira (1995) found that the $(B - V)$ colors of several apparently
unreddened SNe Ia evolved in a nearly identical fashion from
$t$ = 30 to 90 days. This property, which was further discussed
by Phillips et al. (1999), is shown in Figure 7 as the ``L$-$P law"
(which stands for the ``Lira-Phillips law"). Phillips et al. (1999) noted
that individual SNe could display systematic residuals with respect
to this fit and suggested an intrinsic dispersion of $\sim$ 0.05 mag
for the fit, which seems to be too small for the SNe in Figure 7: 
SNe 1991T and 1992al are $\sim$ 0.1 mag redder than the 
fit. While the bad fit to SN 1991T could be caused by an incorrect estimate
of its reddening, SN 1992al is an apparently unreddened SN Ia (Hamuy
et al. 1996b; Phillips et al. 1999). The
most disturbing case, however, is SN 2000cx, which shows a color
$\sim$ 0.2 mag {\it bluer} than the fit. Increasing the
reddening to SN 2000cx will only make it intrinsically bluer, 
further deviating from the fit. Thus, although the L-P law 
provides an independent method to estimate the extinction toward SNe Ia, 
it should be used with caution for any individual SN.

The late-time $(B - V)_0$ evolution ($t >$ 90 days) of SN 2000cx
is well sampled. These points, despite large uncertainties, 
suggest that SN 2000cx has a nearly constant color between $t$ = 90
and 150 days.  SN 1992bc also seems to have a nearly constant color from
$t$ = 80 to 110 days, while SN 1992al seems to have a linear decrease
during this period. Hence, the late-time color evolution of SNe Ia seems
to be diverse, although the current sample is too small to permit
a definitive conclusion. 

The $(V - R)_0$ color evolution of the SNe Ia is shown in Figure 8. 
SN 2000cx has the bluest color until $t$ = 25 days. It has a linear
decline in color at a rate of 0.014 mag day$^{-1}$ between $t$ = $-$8 to 
6 days, then switches to a faster decline at a rate of 0.030 
mag day$^{-1}$ until $t$ = 12 days. This faster decline phase also correlates
with the transition phase in the $B$-band light curve and the plateau
phase in the $(B - V)_0$ color evolution.  The bluest color of SN 2000cx,
$(V - R)_0 \approx -0.4$ mag, occurs at $t \approx$ 12 days. This color is
about 0.2 mag bluer than that of SN 1994D and 0.35 mag bluer
than that of SN 1992bc at the same epoch. There is a dramatic difference
in the $(V - R)_0$ color evolution before $t$ = 20 days between 
SN 2000cx and SN 1992bc: SN 1992bc shows a very flat evolution, while SN 2000cx
develops a prominent ``dip" during this time. It is thus interesting
to notice that these two SNe share similar evolution from $t$ = 20 to 
30 days. At late times ($t >$ 30 days), SN 2000cx shows a very normal
$(V - R)_0$ color evolution compared to the other SNe Ia in Figure 8.

Figure 9 shows that SN 2000cx has the bluest $(V - I)_0$ color 
of all the SNe Ia in the comparison sample (except during $t$ = 20 to 30 days,
at which time SN 1992bc is bluer than SN 2000cx). From $t$ = $-$8 to 
10 days, SN 2000cx becomes progressively bluer by $\sim$ 0.9 mag,
about 0.3 mag more 
than the other SNe do. The $(V - I)_0$ color at $t$ = 10 days
is $-$1.0 mag, the bluest of all known SNe Ia.
During the transition phase in the $B$-band light
curve and the plateau phase in the $(B - V)_0$ color curve ($t$ = 6 to 15
days),  the $(V - I)_0$ color of SN 2000cx develops the bluest ``dip." 
All the SNe in Figure 9 show an approximately linear increase in
color between $t$ = 10 and 30 days, although SN 2000cx has the biggest
slope (about 0.07 mag day$^{-1}$). SN 2000cx also has the bluest $(V - I)_0$ 
color at $t >$ 30 days, during which time all the SNe show
a linear decrease in color. 

It is interesting to note that even though the premaximum brightening 
of SN 2000cx is exactly the same as that of SN 1994D in all $BVRI$ passbands, 
their colors are quite different: SN 2000cx is  redder by 
0.1 mag in $(B - V)_0$, and bluer by 0.1 mag in $(V - R)_0$ and $(V - I)_0$,
than SN 1994D. Thus color evolution of SNe Ia, if available, may reveal 
differences among SNe Ia that are otherwise hidden in their light-curve shapes. 

In summary, the color evolution of SN 2000cx is peculiar. In $(B - V)_0$, the
relatively red color at $t < 6$ days, the plateau phase between $t$ = 6 and 15
days, and the blue late-time tail ($t > 30$ days) are unique. The very blue
color at $t$ = 5 to 20 days in $(V - R)_0$ and $(V - I)_0$, and the blue
late-time tail ($t >$ 30 days) in $(V - I)_0$, are also puzzling. In \S 3, we
will investigate how the peculiarities in the light curves and color curves of
SN 2000cx are correlated with its spectral evolution.

\section{SPECTROSCOPY}

\subsection{General Results}

Spectroscopic observations of SN 2000cx were conducted with the Lick Observatory
1-m Nickel telescope (hereafter L1) using its spectrograph, and with the Lick 3-m Shane 
telescope (hereafter L3) using the Kast spectrograph (Miller \& Stone 1993). The 
journal of observations is given in Table 5.

\begin{deluxetable}{llllccll}
\tablecaption{Journal of spectroscopic observations of SN 2000cx}
\tablehead{
\colhead{UT Date} & \colhead{$t$\tablenotemark{a}} &\colhead{Tel.\tablenotemark{b}}
&\colhead{Range\tablenotemark{c}}&\colhead{Air.\tablenotemark{d}} &
\colhead{Slit} &\colhead{Exp.} &\colhead{Observer(s)\tablenotemark{e}} \\
 & (day) & &{~~~~~(\AA)}&&(arcsec)&{~~(s)~~}&
}
\startdata
2000-07-23 &$-$3  &L1&3801-7340 & 2.0&  2.9 & 4$\times$900 &  EG\\
2000-07-24 &$-$2  &L1&3769-6793 & 2.0&  2.9 & 4$\times$900 &  EG\\
2000-07-25 &$-$1  &L1&3769-7340 & 2.1&  2.9 & 4$\times$900 &  EG\\
2000-07-26 &+0  &L1&3819-7142 & 2.1&  2.9 & 4$\times$900 &  EG\\
2000-07-27 &+1  &L1&3769-7241 & 2.1&  2.9 & 4$\times$900 &  EG\\
2000-07-28 &+2  &L3&3323-9919 & 1.7&  2.0 & 2$\times$200 &  AF, RC\\
           &   &L3p&4300-6900 & 1.4&  3.0 & 4$\times$1500&  AF, RC\\
2000-08-01 &+6  &L3&3323-10193& 1.1&  2.0 & 2$\times$300 &  MR\\
2000-08-02 &+7  &L3&3323-10400& 1.1&  2.0 & 2$\times$300 &  MR\\
2000-08-03 &+8  &L1&3918-7390 & 2.1&  2.9 & 3$\times$1200&  EG\\
2000-08-05 &+10 &L1&3918-7390 & 2.1&  2.9 & 3$\times$1200&  EG\\
2000-08-07 &+12 &L1&3918-7390 & 2.1&  2.9 & 3$\times$1200&  EG\\
2000-08-10 &+15 &L1&3918-7390 & 2.2&  2.9 & 3$\times$1200&  EG\\
2000-08-15 &+20 &L1&3918-7390 & 2.0&  2.9 & 3$\times$1200&  EG\\
2000-08-18 &+23 &L1&3918-7390 & 2.2&  2.9 & 3$\times$1200&  EG\\
2000-08-20 &+25 &L1&3918-7390 & 1.9&  2.9 & 3$\times$1200&  EG\\
2000-08-22 &+27 &L1&3918-7390 & 2.1&  2.9 & 3$\times$1200&  EG\\
2000-08-24 &+29 &L1&3918-7390 & 1.8&  2.9 & 3$\times$1200&  EG\\
2000-08-26 &+31 &L1&3918-7390 & 1.7&  2.9 & 3$\times$1200&  EG\\
           &+31 &L3p&4300-6900 & 1.4&  3.0 & 2$\times$1500& DL, WL  \\
           & && & &   & 2$\times$1800&  \\
2000-08-27 &+32 &L3&3273-10365& 1.1&  2.0 & 450   &  AF, WL, DL\\
           &+32 &L3p&4300-6900 & 1.4&  3.0 & 4$\times$1500& AF, WL, DL\\
2000-09-06 &+42 &L3&3273-10365& 1.1&  2.0 & 700   &  AF, RC, WL\\
2000-09-26 &+62 &L3&3273-10316& 1.1&  2.0 & 1200  &  AF, RC, WL\\
2000-10-06 &+72 &L3&3273-10316& 1.1&  2.0 & 1800  &  AF, RC\\
2000-10-24 &+90 &L3&3273-10316& 1.1&  2.0 & 1800  &  AF, RC\\
2000-11-01 &+98 &L3&3273-9907 & 1.2&  2.0 & 2$\times$1800&  AF, WL\\
2000-11-29 &+126&L3&3273-10380& 1.1&  2.0 & 1800  &  AF, RC\\
2000-12-21 &+148&L3&3273-10415& 1.1&  2.0 & 2700  &  AF, RC, MM\\
\enddata
\tablenotetext{a}{Days since $B$ maximum brightness (JD = 2451572.2), rounded to the nearest day.}
\tablenotetext{b}{L1 = Lick 1-m/Nickel reflector + spectrograph; L3(p) = 
Lick 3-m/Shane reflector + Kast double spectrograph (``p" denotes polarimeter attached).}
\tablenotetext{c}{Observed wavelength range of spectrum. }
\tablenotetext{d}{Average airmass of observations.}
\tablenotetext{e}{AF = Alex Filippenko; DL = Douglas Leonard;  EG = Elinor Gates; MM = Maryam Modjaz; 
MR = Michael Rich; RC = Ryan Chornock; WL = Weidong Li.}
\end{deluxetable}

All one-dimensional sky-subtracted spectra were extracted optimally
(Horne 1986) in the usual manner, generally with a width of $\sim 10\arcsec$
along the slit. Each spectrum was wavelength and flux calibrated,  and 
corrected for continuum atmospheric extinction and telluric absorption
bands (Bessell 1999; Matheson 2000). 

In general, the position angle of the slit was aligned along the parallactic
angle for the L3 observations (except those in spectropolarimetric mode), so
that the spectral shape does not suffer from differential light loss
(Filippenko 1982).  The L1 observations, on the other hand, were all obtained
with a fixed angle of $0^\circ$ at relatively large airmasses (1.7 to 2.2),
making their continuum shape unreliable.  Leonard (2001, his Figure 5.5) also
found that the continuum shape at $\lambda \geq$ 5800 \AA\, of some of the L1
spectra of SN 1999em suffered from a mysterious problem, whose symptom is that
successive observations during the same night on a few nights showed large
variations. This problem, however, is not present in the L1 data on SN 2000cx
-- the reductions show that successive observations are consistent with each
other. Li et al. (2001, in preparation) show that the L1 data on SN Ia 1999by
do not suffer from this problem either.  Thus, the problem found by Leonard
(2001) appears to be sporadic.

Besides normal spectroscopy, spectropolarimetry (range 4300 to 6900 \AA) of SN
2000cx was obtained with L3 on three nights: 2000 July 28, August 26, and
August 27. Preliminary analysis of the observations on 2000 July 28 (Leonard et
al. 2001) indicates a continuum polarization of about 0.5\%. A strong
polarization modulation (about 0.3\%) across the Si~II $\lambda$6355 line is
found, implying significant polarization intrinsic to SN 2000cx itself,
possibly resulting from an aspherical scattering atmosphere (e.g., H\"oflich
1991). Detailed results of the spectropolarimetry will be the topic of a future
paper; here only the total-flux spectra are reported.

The spectral evolution of SN 2000cx from $t$ = $-$3 to 148 days is shown
in Figures 10 and 11. All spectra shown in this paper have been corrected 
for reddening and for the host-galaxy redshift. For SN 2000cx, we adopt a reddening
of $E(B - V) = 0.08$ mag as discussed earlier, and a redshift of 2421
km s$^{-1}$ from NED\footnote{NED (NASA/IPAC Extragalactic Database) is
operated by the Jet Propulsion Laboratory, California Institute of Technology,
under contract with the National Aeronautics and Space Administration.}.
With 29 spectra obtained from $t$ = $-$3 to +148 days, SN 2000cx is 
one of the spectroscopically best-observed SNe Ia.

Our strategy for studying the spectral evolution of SN 2000cx is to use Figures
10 and 11 as a guide, and to conduct detailed comparisons between SN 2000cx and
other SNe Ia at different epochs in Figures 12 through 17.  The example of a
normal SN Ia used in the comparison is SN 1994D (Patat et al. 1996; Filippenko
1997). Since the premaximum spectra of SN 2000cx are similar to those of SN
1991T, we include in the comparison SN 1991T (Filippenko et al. 1992b) and SN
1997br, another SN 1991T-like SN Ia (Li et al. 1999).  The line identifications
adopted here are taken from Kirshner et al. (1993), Jeffery et al. (1992), and
Mazzali, Danziger, \& Turatto (1995).

\subsection{The First Three Weeks}

The top spectrum in Figure 10 was obtained with L1 on 2000 July 23, 6 days 
after the discovery of SN 2000cx. Figure 12 shows a comparison of this
spectrum with those of other SNe Ia at similar epochs ($t$ = $-$3 days).
The spectrum of SN 2000cx is similar to that
of SN 1997br:  Fe~III $\lambda$4404 and Fe~III $\lambda$5129 are the two
major absorption lines, and the Si~II $\lambda$6355 line is much weaker than
that of SN 1994D, with an asymmetric profile  similar to that seen in 
SN 1997br. There seems to be some high-excitation Si~III lines  in
the spectrum; Si~III $\lambda$4560 is weak but apparent, and the 
line at $\sim$ 5500 \AA\, marked with a ``?" may be Si~III $\lambda$5740.

The overall continuum of SN 2000cx seems to be redder than
those of SNe 1994D and 1997br. While this could be caused by the
differential light loss for the L1 observations, it is
confirmed by the concurrent red photometric $(B - V)_0$ values. A redder continuum,
with a lower color temperature, 
is often regarded as being associated with a lower radiation temperature as well.
Pinto \& Eastman (2000), however,
demonstrate that the observed spectral continuum shape (and thus 
the color temperature) may have less to do with close thermal
coupling between the gas and the radiation field and more to do 
with the distribution of lines and branching ratios in the complex
atomic physics of the iron group. In other words, the redder continuum
of SN 2000cx does not necessarily mean that it has a cooler 
radiation field -- it may be caused by different excitation conditions
and/or distribution of the iron-peak elements. As discussed later, 
we favor the latter explanation. 

One interesting feature of the SN 2000cx spectrum at $t$ = $-$3 days 
is the narrow emission at about 4060 \AA\, (marked
by a ``?" in Figure 11), which appears in the spectra of SN 2000cx until
$t$ = 7 days. It is often accompanied by another narrow emission line at
about 3860 \AA\, (see the $t$ = 2, 6, and 7 days spectra
in Figure 10; also see Figure 12). The 
nature of those two narrow emission lines is not clear. While their
wavelengths are coincident with those of H$\delta$ and H$\zeta$ blueshifted
by about 3,000 km s$^{-1}$, the lack of other hydrogen Balmer lines makes
these identifications very unlikely. 
Perhaps the apparent ``emission lines" are explained by a lack of absorption at these
wavelengths.

From $t$ = $-3$ to +2 days (Figure 10),  Si II $\lambda$6355 
becomes progressively
stronger.  Fe III $\lambda$4404 remains prominent, although its
profile shows evolution  during this time -- the blue wing becomes 
steeper from $t$ = $-$3 to $-$1 day, then flatter again to $t$ = 
2 days, while Si III $\lambda$4560 at the red wing becomes weaker
and vanishes at $t \approx$ 2 days. The blue wing
of Fe III $\lambda$5129  seems to be contaminated
by a strengthening absorption that is probably Si II $\lambda$5051. 
The S II $\lambda$5468 and S II $\lambda\lambda$5612, 5654 lines also 
strengthen with time.

By $t$ = 2 days, the spectrum of SN 2000cx looks
almost normal, though a detailed comparison (Figure 13) shows that there
are still noticeable differences. Neither SN 1991T nor SN 1997br have
spectra at this epoch so they are not included in the comparison. 
The largest difference between SN 2000cx and 
SN 1994D at this epoch is the relatively weaker features for the 
intermediate-mass elements (Si II, Ca II, S II, and O I), and
the relatively stronger Fe III $\lambda$4404 and Fe III $\lambda$5129
lines,  in the spectrum of SN 2000cx. The two mysterious narrow emission 
lines discussed earlier are still present. There are
also apparent differences in the spectral range from 7800 to 8400 \AA\, --
SN 1994D shows a strong Ca II IR triplet, while SN 2000cx shows
four absorption wiggles.  The two wiggles at the red side correspond to the 
Ca II IR triplet, but the two at the blue side do not have counterparts
in the spectrum of SN 1994D. We have difficulty identifying these two
wiggles at the blue side, but suspect that they may be part of the Ca II 
IR triplet absorption caused by some unique distribution of Ca in the ejecta
of SN 2000cx. This identification, however, requires part of the 
Ca in the ejecta of SN 2000cx to have very high expansion velocity
 ($\geq$ 28,000 km s$^{-1}$). Hatano et al. (1999) found evidence for 
very high expansion velocity for Ca II features in the spectra of SN 1994D,
 but at a much earlier phase ($t$ = $-$8 days). They suggested that the 
lower-velocity matter represents freshly synthesized material, while the
highest-velocity matter is likely to be primordial.  Similar wiggles can be seen in the
 near-maximum spectrum 
of SN 1991T (Filippenko et al. 1992b), possibly in the data on SN 1997br
(Li et al. 1999),  and in other SN 1991T-like objects
(Li et al. 2001, in preparation), but not in normal SNe Ia such as 
SN 1981B (Branch et al. 1983), SN 1989B (Wells et al. 1994), and
 SN 1992A (Kirshner et al. 1993).
It is unclear how the differences in the evolution of Ca~II profiles 
are related to the models of SNe Ia. 

Between $t$ = 6 and 15 days, SN 2000cx experienced the transition phase in 
the $B$-band light curve, the plateau phase in the $(B - V)_0$ color curve,
and the bluest color in the $(V - R)_0$ and $(V - I)_0$ evolution. 
Spectroscopically (Figure 10), it is puzzling how the $(B - V)_0$ color of 
SN 2000cx could be nearly constant during this period because there
is apparent evolution in the spectra of SN 2000cx that corresponds to the 
broad $BV$ passbands (3500 -- 6500 \AA), especially during $t$ = 10 to 
15 days. The S~II $\lambda$5468 and
S~II $\lambda\lambda$5612, 5654 lines become progressively weaker 
and vanish at $t \approx$ 12 days. The Na I D and Si~II $\lambda$5972
blend at $\sim$ 5750 \AA\, becomes much stronger and dominates the spectrum
at $t$ = 15 days. The Fe~III $\lambda$5129 line is also contaminated by
Si~II $\lambda$5051. One possible explanation for the constant $(B - V)_0$
color is that the change in the features is compensated by a change 
in the continuum shape. While most changes in features happened in the $V$ band
(4800 -- 6500 \AA) and they tend to increase the $(B - V)_0$ color (because
of fewer absorption lines in the $V$ band), the continuum shape seems to become bluer
from $t$ = 10 to 15 days which reduces the $(B - V)_0$ color. These two
opposite trends cancel out and result in a constant color in $(B - V)_0$. 
The blue continuum is also consistent with the very blue $(V - R)_0$ and
$(V - I)_0$ color evolution during this time. Since the spectral features
in the $R$ band (5700 -- 7500 \AA) do not change significantly, the blue 
 $(V - R)_0$ color is almost certainly caused by the blue continuum 
of the spectra. Unfortunately, our spectra do not cover the whole $I$ band
during this time, so it is unclear how the spectral features (in particular,
the Ca~II IR triplet) change in this spectral range (7300 -- 9000 \AA).

Also unique to SN 2000cx at this time is the
persistence of  strong, high-excitation Fe~III and weak Fe~II lines. The
high-excitation Fe~III lines are usually seen in the premaximum spectra of 
SN 1991T-like objects, but in the case of SN 2000cx, th Fe~III $\lambda$4404
is very strong at 4260~\AA\, even in the $t$ = 15 day spectrum (Figure 14).
While it could be argued
that this line may be the Fe~II $\lambda$4555 line at high expansion
velocities, we dispute this identification for two reasons: (1) the good temporal
coverage of the spectra of SN 2000cx from $t$ = $-$3 to 15 days shows clearly
that this line evolved from the Fe~III $\lambda$4404 line in the $t$ = $-$3 day
spectrum; and (2) the Fe~II $\lambda$4555 line is seen to develop in the $t$ =
20 day spectrum (Figure 15) at a more reasonable wavelength. A high expansion velocity
of more than 20,000 km s$^{-1}$ is required to identify the absorption
at 4260 \AA\, as the Fe~II $\lambda$4555 line,  while the expansion velocity
measured from the Fe~II $\lambda$4555 line at $t$ = 20 days is only about
10,000 km s$^{-1}$. It is unlikely that the expansion velocity measured
from the same line could change by more than 10,000 km s$^{-1}$ in just
5 days. 

Other evidence for strong Fe~III and weak Fe~II lines in the spectra
of SN 2000cx is the presence of Fe III $\lambda$5129  and the 
absence of Fe II $\lambda$5169 in the $t$ = 15 day spectrum (Figure 14).
The Fe II lines that usually contaminate
Si II $\lambda$6355 are also absent, even in the $t$ = 20 day spectrum
(Figure 15). Moreover, SN 2000cx lacks the Fe~II $\lambda$5535 absorption
in the $t$ = 20 day spectrum. 

The slow change in the ionization stages of Fe is consistent with 
the prominent ``dip" and faint secondary peak evolution for 
the $I$-band light curve of SN 2000cx, as discussed in \S 2.

The evolution of the Si II lines in the spectra of SN 2000cx is also
peculiar during this time. Si II $\lambda$6355 
is strong and keeps a P-Cygni profile until $t$ = 20 days (Figure 15).
This is different from most other SNe Ia,
whose Si~II $\lambda$6355 line usually shows progressive contamination
by Fe~II lines at the blue and the red wings. 
Figure 15 demonstrates that Fe~II lines totally replaced 
Si~II $\lambda$6355 in SN 1997br, while they seriously contaminated the 
blue and the red wings of the Si~II $\lambda$6355 line in SN 1994D. 
While the P-Cygni profile of Si~II $\lambda$6355 in SN 2000cx
could be explained by the weakness of the Fe~II lines,
there is other evidence for strong Si~II lines in the spectrum.
The Si~II $\lambda$5972 and Na I D blend in SN 2000cx is much stronger
than that in SN 1994D and SN 1997br (Figures 14 and 15). The Si~II $\lambda$5051 line
is also prominent. In fact, because of the strength of this Si~II line,
and the weak Fe~II $\lambda$5108 line, the spectral range between 
4600 and 5200 \AA\, forms a broad absorption trough in the $t$ = 20 day spectrum
of SN 2000cx (Figure 15), in direct contrast to the
two emission lines in the spectra of SNe 1994D and 1997br. 

\subsection{Entering the Nebular Phase}

The most important spectral evolution of SN 2000cx during $t$ = 23
to 32 days (Figures 10 and 11) is the emergence of the Fe~II lines and the
weakening of the Si II lines. Fe~II $\lambda$4555 appears
at $t$ = 20 days and develops into a P-Cygni profile. The 
Fe II $\lambda$5169 line also replaces Fe III $\lambda$5129. 
The Fe II lines at the blue and the red wings of Si II $\lambda$6355
appear and strengthen over the time. The Si II lines, on the
other hand, become progressively weaker. Si II $\lambda$6355 
almost disappears in the $t$ = 32 day spectrum, while the Si II
$\lambda$5972 plus Na I D blend at 5750 \AA\, becomes progressively
weaker, most likely due to the decline of Si II $\lambda$5972.
The weakening of Si II $\lambda$5051 and the strengthening
of Fe II enable the two 
emission lines to appear between 4600 \AA\, and 5200 \AA, similar to those of 
other SNe 
Ia. By $t$ = 32 days (Figure
16), the spectrum of SN 2000cx looks very similar to that of SN 1994D,
although there are still subtle differences in the line intensities. 

In the $t$ = 32 day spectrum of SN 2000cx (Figure 16), 
there seems to be additional absorption at the blue side of 
Ca II H \& K, which might be caused by Co II $\lambda$3755,
or Ti II $\lambda$3760, or a high expansion velocity component of 
Ca II H \& K.
The identification of Si II $\lambda$3858 is less plausible, 
since all the other Si II lines are weak in the spectrum and become
weaker, and this feature seems to strengthen from $t$ = 32 to 42 days.
This line weakens after
$t$ = 42 days and vanishes at $t \approx$ 98 days (Figure 11). 
The evolution of this feature appears unique to SN 2000cx -- we could not  find 
other SNe Ia that have a similar line, either in the literature or 
in our own spectral database. Understanding the evolution of this line
through detailed spectral synthesis may provide clues to 
the nature of SN 2000cx. 

The Ca II IR triplet
absorption is stronger in SN 2000cx than in SNe 1994D and 1997br. 
Inclusion of the absorption marked ``?" in Figure 16 as part of 
the Ca II IR triplet line may strengthen our earlier suggestion
that the two blue wiggles in Figure 13 are caused by Ca, 
although our spectra do not show clearly how those
wiggles evolved.   

One explanation of the strengthening of the Fe II lines and the weakening
of the Si II lines during this period is that the ejecta of SN 2000cx have expanded and 
become optically thin for the intermediate-mass elements, and the 
photosphere has receded into the region with more iron-peak elements.
Alternatively, the temperature of the ejecta may have dropped enough to
make most iron-peak materials  singly ionized.  
We think both of these causes may contribute to the spectral appearance
of SN 2000cx during this period. 

After $t$ = 32 days the spectral evolution of SN 2000cx is relatively slow,
and not very different from that of normal SNe Ia.
Apart from the absorption at the blue side of Ca~II H \& K,
the other major
change is the development of strong absorption lines in the
7500 -- 9000 \AA\, region. These 
are caused by a combination of O I $\lambda$7773, the Ca~II IR 
triplet, and O I $\lambda$9264. Figure 17 shows a comparison 
of the late-time ($t > 100$ days) spectra of SNe 1991T, 1994D, and 2000cx,
with SN 2000cx having the strongest absorption in the red. 
The development of these absorption lines might cause the 
fast decline in the $I$-band photometry at $t > $30 days.
Although spectra of SNe Ia at this stage do not
have a well-defined continuum, the SN 2000cx spectrum in Figure 17 still
seems to be the bluest of the three,  consistent with the late-time
color evolution of SN 2000cx.

\subsection{Expansion Velocities}

The expansion velocities ($V_{exp}$) as inferred from observed minima
of absorption lines in the spectra 
may provide some clue to the nature of SN Ia explosions (Branch, Drucker,
\& Jeffery 1988). Figure 18 shows the expansion velocities derived from 
several lines.  The Fe III lines of SN 2000cx
not only last the longest, but also yield the highest expansion velocities. The $V_{exp}$
derived from the Fe III $\lambda$4404 line is higher by about 1,000 km s$^{-1}$
for SN 2000cx than for SN 1991T, while it is higher by about
2,000 km s$^{-1}$ from the Fe III $\lambda$5129 line.  Fe II
$\lambda$4555 also yields a $V_{exp}$ for SN 2000cx that
is about 1,000 km s$^{-1}$ higher than those for SNe 1991T and 1994D.
The $V_{exp}$ derived from the Fe III $\lambda$5129 line seem to 
be systematically higher by about 2,000 km s$^{-1}$ than those
from the Fe III $\lambda$4404 line, which is probably caused by
the blending of Fe III $\lambda$5129 with Si II $\lambda$5051.

The S II $\lambda$5468 line does not always have a well-defined profile,
so the corresponding  $V_{exp}$ measurements for SN 2000cx have a large scatter.
Nevertheless, they are systematically higher by about 2,000 to 3,000 
km s$^{-1}$ than those for SNe 1991T (only one point) and 1994D. The 
values of $V_{exp}$ from S II $\lambda\lambda$5612, 5654 for SN 2000cx
are quite self-consistent
and  are higher by about 2,000 km s$^{-1}$ than those of SNe 1991T (only
one point) and 1994D. 

The $V_{exp}$ measurements from Si II $\lambda$6355 for SN 2000cx
show a unique flat evolution -- they remain nearly constant at 
12,000 km s$^{-1}$ from $t$ = $-$3 to 40 days,  while all the other SNe Ia
in Figure 17 exhibit a gradual decline in $V_{exp}$. 
Moreover, the expansion velocities for SN 2000cx are the highest
among all the SNe Ia in the comparison sample. These
facts perhaps suggest that the Si zone in the expanding ejecta of 
SN 2000cx was confined to a fairly restricted layer characterized by
velocities around 12,000 km s$^{-1}$. 

The higher expansion velocities of SN 2000cx 
imply more kinetic energy  if the mass of its ejecta is 
similar to that of other SNe Ia.  This 
implication will be used in our discussions of the  theoretical models
for SN 2000cx.

In summary, the spectral evolution of SN 2000cx is unique. 
The change of the excitation stage for Fe is unusually slow  --  
the Fe III lines are present in the spectra until $t \approx$ 15 days, 
while the Fe II lines are weak in the spectra until $t \approx$ 23 days. 
The Si II lines are weak in the premaximum spectra, but strengthen
over time until $t \approx$ 7 days, and stay relatively strong 
until $t \approx$ 20 days. The evolution of the Ca II IR triplet
is also peculiar, showing wiggles in the spectra near maximum  brightness 
and strong absorption at late times. SN 2000cx also has  
high expansion velocities. 

\section{DISCUSSION}

\subsection{Absolute Magnitudes}

\subsubsection{Absolute Magnitudes from Distance Estimates}

The host galaxy of SN 2000cx, NGC 524, is the central galaxy in the compact
group CfA 13 of Geller \& Huchra (1983). 
The mean heliocentric radial velocity of the eight
known group members is 2476 km s$^{-1}$, with a line-of-sight
dispersion $\sigma_{V} \approx$ 205 km s$^{-1}$. The heliocentric velocity
of NGC 524 itself (2421 km s$^{-1}$ from NED) is just 55 km s$^{-1}$
smaller, i.e., within $\sim (1/4) \sigma_{V}$ of the group mean.
In addition, NGC 524 is at (or close to) the projected geometric center of 
the group, and is almost as luminous as all the other group 
members combined. In short, NGC 524 is a giant galaxy dominating its
surrounding cluster of galaxies. 

Based on results from the 1.2 Jy {\it IRAS} survey (assuming
$\beta$ = 0.5 in a linear perturbation theory), M. Davis (2001, private
communication) concludes that NGC 524 does not participate in the 
local bulk flow of the Local Group region, but is infalling to the 
Perseus-Pisces supercluster, and has a predicted outflow of 
418$\pm$200 km s$^{-1}$ in the Local Group frame. The corrected recession
velocity of NGC 524 in the Local Group frame is 2192$\pm$200 km s$^{-1}$.
Thus, we find a distance
of 34$\pm$3 ($H_0$/65) Mpc and a distance modulus of $\mu = (m - M) = 
32.64(\pm0.20) - 5~$log$(H_0/65)$ mag. 

\begin{deluxetable}{lccccc}
\tablecaption{Absolute peak magnitudes of SN 2000cx\tablenotemark{a}}
\tablehead{
\colhead{Method} &\colhead{$\mu$} & \colhead{$M_B$} & \colhead{$M_V$} &
\colhead{$M_R$} & \colhead{$M_I$} }
\startdata
\\
\multicolumn{6}{c}{From distance estimates}\\
\\
\hline

$v_{CMB}$ ($H_0 = 58.5$) & 32.87$\pm$0.21& $-$19.77$\pm$0.41 & $-$19.87$\pm$0.21 & $-$19.70$\pm$0.21 & $-$19.38$\pm$0.21 \\
\hline
\\
\multicolumn{6}{c}{From light-curve fits}\\
\\
\hline
MLCS\tablenotemark{b} & 32.53$\pm$0.35& $-$19.61$\pm$0.21 & $-$19.59$\pm$0.21 & $-$ & $-$19.31$\pm$0.22 \\
Stretch ($-$8 to 32d) & $-$& $-$19.43$\pm$0.20 & $-$19.43$\pm$0.17 & $-$ & $-$19.20$\pm$0.18 \\
Stretch ($-$8 to 1d) & $-$& $-$19.31$\pm$0.12 & $-$19.35$\pm$0.10 & $-$ & $-$19.21$\pm$0.14 \\
Stretch (\,\,\,\,1 to 32d) & $-$& $-$19.72$\pm$0.14 &$-$19.68$\pm$0.12 & $-$ & $-$19.35$\pm$0.15 \\
$\Delta m_{15}(B)$ & $-$ & $-$19.70$\pm$0.14 & $-$19.66$\pm$0.12 & $-$ & $-$19.34$\pm$0.15 \\
Two$-$params\tablenotemark{c} &$-$& $-$19.48$\pm$0.14 & $-$19.54$\pm$0.10 & $-$  & $-$19.24$\pm$0.12 \\
\hline
\enddata
\tablenotetext{a} {See text for details.}
\tablenotetext{b} {The absolute magnitudes for a normal SN Ia are adopted from 
Parodi et al. (2000).}

\tablenotetext{c}{The two-parameter luminosity correction method. The
luminosity corrections are derived from the parameterization of Drenkhahn \&
Richtler (1999), and the absolute magnitudes for a normal SN Ia are adopted
from Parodi et al. (2000).}
\end{deluxetable}

Using the apparent peak magnitudes listed in Table 4, together with
our estimates of the extinction and distance, we derive the peak absolute
magnitudes of SN 2000cx in all filters in Table 6 (under category
``from distance estimates"). We adopt $H_0$ from
Parodi et al. (2000; 58.5 km s$^{-1}$ Mpc$^{-1}$), but do not include
the uncertainty in $H_0$ in the error analysis. 
Compared to the absolute magnitudes for a normal SN Ia from Parodi 
et al. (2000), those of SN 2000cx are overluminous  
by 0.22, 0.34, and 0.13 mag in $M_B, M_V, M_I$, respectively. 

\subsubsection{Absolute Magnitudes from Light-Curve Fits}

In earlier discussions we speculated that the peculiar photometric
evolution of SN 2000cx might challenge the light-curve fitting
techniques for SNe Ia. Here we present more details.

There is a fairly large scatter among the published absolute magnitudes for
a normal SN Ia (e.g., Vaughan et al. 1995; 
Riess, Press, \& Kirshner 1996; Saha et al. 1999;
Phillips et al. 1999; Parodi et al. 2000), most of which 
is caused by the different methods used to estimate the reddenings of
the SNe Ia. 
To facilitate discussion, throughout this paper we adopt the
absolute magnitudes for a normal SN Ia from the 
weighted mean of eight SNe Ia calibrated 
through Cepheid distances of their parent galaxies (Parodi et al. 2000): 
$M_B = -19.55\pm0.07$ mag, $M_V = -19.53\pm0.06$ mag,
and $M_I = -19.25\pm0.09$ mag. {\it The light-curve fitting techniques
are only used to derive the luminosity corrections to these absolute
magnitudes.} The weighted mean 
$\Delta m_{15}(B)$ of the eight calibrated SNe Ia is 1.08$\pm$0.02 mag. 
Several light-curve fitting techniques correct the absolute 
magnitudes of SNe Ia to a ``standard SN Ia" with $\Delta m_{15}(B) = 1.10$
mag, and we do not attempt to reconcile the small difference (0.02 mag) in
those two $\Delta m_{15}(B)$ values. The $R$-band absolute magnitude
is not well studied in the SNe Ia calibrated with Cepheid variables, and 
it will not be considered in most of our discussions.

Phillips et al. (1999) used the ``Lira-Phillips law" to study the 
reddening of a sample of 62 SNe Ia, and revised the absolute 
magnitude vs. $\Delta m_{15}(B)$ relation in a quadratic form.
Using the measured $\Delta m_{15}(B)$ of SN 2000cx 
(0.93$\pm$0.04 mag) and equations 20 through 22 in Phillips
et al. (1999), we derive the luminosity corrections for 
SN 2000cx when it is adjusted to a standard decline rate of 
$\Delta m_{15}(B)$ = 1.1 mag: $\Delta M_B^{cor} = -0.15\pm0.08$ mag,
$\Delta M_V^{cor} = -0.13\pm0.08$ mag, and $\Delta M_I^{cor} = -0.09\pm0.07$ mag. 
These numbers indicate that SN 2000cx is slightly overluminous compared 
with a normal SN Ia with $\Delta m_{15}(B) = 1.1$ mag. Using the 
absolute magnitudes of a normal SN Ia from Parodi et al. (2000), we derive the 
absolute magnitudes of SN 2000cx to be $M_B = -19.70\pm0.14$ mag, 
$M_V = -19.66\pm0.12$ mag, and $M_I = -19.34\pm0.15$ mag.

We have done a MLCS fit to SN 2000cx and the results are shown in the
left panel of Figure 19. 
This is the worst MLCS fit we have ever seen for a SN Ia. While the $V$ and
$R$ fits may be marginal, the $B$ and $I$ fits are poor. 
The model $B$-band light curve peaks about two days earlier than the
observed one, and the slow decline in the observations is
not reproduced. In the $I$ band the deep decline after the first
peak is not reproduced by the model. It is clear that the photometric
evolution of SN 2000cx does not conform to the family of SNe Ia
constructed in MLCS. 

The MLCS fit reports a distance modulus to SN 2000cx
of $\mu$ = 32.53$\pm$0.35 mag,
and a ``luminosity correction" $\Delta$ = $-$0.06 $\pm$ 0.20 mag. This distance
modulus is slightly smaller than, but within the uncertainties of,  the ones estimated
from the recession velocity of NGC 524. The derived absolute magnitudes
of SN 2000cx are $M_B = -19.61\pm0.21$ mag, $M_V = -19.59\pm0.21$ mag,
and $M_I = -19.31\pm0.22$ mag if the absolute magnitudes of a normal 
SN Ia from Parodi et al. (2000) are 
used. Considering the poor MLCS fit, however, these values 
may not be representative of SN 2000cx. 

The right panel of Figure 19 shows three fits to the peak of the $B$-band light curve
of SN 2000cx ($t$ = $-$8 to 32 days) using the ``stretch method" (Perlmutter 
et al. 1997).
The normalized flux for the $B$-band template is adopted from the
``Parab-18" model of Goldhaber et al. (2001).  The solid, dash-dotted, and 
dashed lines are the fits for all the data points ($t$ = $-$8 to 32 days),
the premaximum data points ($t$ = $-$8 to 1 day), and the postmaximum
data points ($t$ = 1 to 32 days), respectively. The stretch factors ($s$) derived
are 0.89$\pm$0.06 (solid line), 0.76$\pm$0.01 (dash-dotted line), and
1.09$\pm$0.02 (dashed line), and the corresponding $\Delta m_{15}(B)$
values are 1.27$\pm$0.18 mag, 1.64$\pm$0.02 mag, and 0.91$\pm$0.03 mag, 
respectively (measured from the fitted curve). We note that the 
solid line (the fit to all the data points from $t$ = $-$8 to 32 days) 
does not reproduce the observations well: it rises a little bit
too slowly, and declines too fast between $t$ = 10 and 30 days. 
The dash-dotted ($s$ = 0.76$\pm$0.01) and the dashed ($s = 1.09\pm0.02$)
lines fit the observations
very well, but yield very different stretch factors. SN 2000cx is clearly
a counterexample to the results of Goldhaber et al. (2001), who claim 
that a single stretch factor applies equally well to the rising and declining
parts of the light curve. 

Using the $\Delta m_{15}(B)$ value measured from the fitted curves and the 
prescription of Phillips et al. (1999), we derive the absolute magnitudes
for SN 2000cx from the stretch method and present them in Table 6, together
with results from the absolute magnitude vs. $\Delta m_{15}(B)$ relation
and MLCS (under category ``from light-curve fits"). The stretch method
yields quite different absolute magnitudes for SN 2000cx when different parts
of the light curve are used in the fit. For example, the fit to the 
premaximum observations indicates that SN 2000cx is subluminous,
while the fit to the postmaximum observations suggests an overluminous event.

One notices that the mismatch between the fit (from both
MLCS and the stretch method) and the data is revealed by the
excellent temporal
coverage of the SN 2000cx observations. If the observations were more sparse,
a better fit could be achieved and the mismatch would not be noticed.
Since most of the observed SNe Ia generally do not have such dense 
temporal coverage, 
it is important to understand the nature of objects like
SN 2000cx and to investigate how frequent they are, so as to assess the
contamination of the observed SN Ia sample (at both low
and high redshifts) by these objects (although there are about a dozen
normal SNe Ia with sampling like SN 2000cx that are fit well by MLCS). 

A two-parameter luminosity correction method was developed recently 
(e.g., Tripp \& Branch 1999; Drenkhahn \& Richtler 1999).
This method corrects the absolute magnitudes of a SN Ia according
to both its $\Delta m_{15}(B)$ and its color ($B_{max} - V_{max}$), 
where $B_{max}$ and $V_{max}$ are the maximum apparent $B$ and $V$ 
magnitudes of the SN {\it corrected only for the Galactic reddening.}
Using the parameterization of Drenkhahn \& Richtler (1999) for the 
luminosity corrections [to a standard SN Ia with $\Delta m_{15}(B) =
1.10$ mag] and our adopted absolute magnitudes of a 
normal SN Ia from Parodi et al. (2000), we derive the following absolute magnitudes for
SN 2000cx: $M_B = -19.48\pm0.14$ mag, $M_V = -19.54\pm0.10$ mag,
and $M_I = -19.24\pm0.12$ mag. These values are listed
in Table 6 as entry ``two-params."

There is one caveat in the two-parameter luminosity correction
method: it does not differentiate between the host-galaxy
reddening and the intrinsic color of SNe Ia. In other words, 
a red ($B_{max} - V_{max}$) color for a SN Ia may be caused by the host-galaxy
reddening to the SN, or the SN may be intrinsically red. Since we cannot
expect the correction coefficients to be the same for the host-galaxy
reddening (which should have coefficients based on the reddening law of 
dust) and the intrinsic color (which should
have coefficients determined from the physics of SNe Ia), 
it may be difficult or impossible to apply 
the method to an individual SN Ia. A good indication of the effect of 
neglecting the 
difference between the host-galaxy reddening and the intrinsic color
of SNe Ia is provided by  the derived coefficients for
($B_{max} - V_{max})$: only 2.1 to 2.5 in the corrections for $M_B$,
much smaller than  the coefficient used in the canonical
Galactic reddening law (4.1). A better approach is to differentiate 
between the host-galaxy reddening and the intrinsic color of SNe Ia, and
use different coefficients for the corrections. 
However, there is no
excellent, general method for estimating the host-galaxy reddening to SNe Ia.
For example, the ``Lira-Phillips law"  (Phillips et al. 1999), though
easy to implement, may suffer from uncertainties larger than initially
estimated, as discussed earlier in this paper.

In summary, the light-curve fitting techniques currently available (MLCS
and the stretch method) fail to reproduce the photometric behavior
of SN 2000cx. The 
absolute magnitude vs. $\Delta m_{15}(B)$ relation and the two-parameter
luminosity correction method use only information derived from  
light curves of SNe Ia and do not attempt to fit them, so it is 
difficult to assess the accuracy of the values derived from these
two methods. 
In the next section, we will estimate the absolute
magnitudes of SN 2000cx from theoretical model considerations 
and compare them to those measured here.

Surprisingly, Table 6 indicates that despite  all the difficulties in
fitting the light curves of SN 2000cx, different techniques yield results 
with reasonably small scatter. The weighted mean absolute magnitudes of 
SN 2000cx are $M_B = -19.53\pm0.15$ mag, 
$M_V = -19.53\pm0.13$ mag, and $M_I = -19.27\pm0.09$ mag, which are 
very similar to the values for normal SNe Ia in Parodi et al. (2000).

\subsection{Theoretical Models for SN 2000cx}

SNe Ia are widely believed to be  thermonuclear explosions of  
mass-accreting carbon-oxygen (C-O) white dwarfs (WDs). The conversion of carbon and
oxygen to iron-peak and intermediate-mass elements releases a large amount 
of energy initially, thereby disrupting the star, and subsequently the radioactive decay of 
$^{56}$Ni to $^{56}$Fe powers the supernova through $\gamma$-ray
photons and positrons. Despite more than half
a century of observations and theoretical studies, however, an 
understanding of the detailed physics leading to SN Ia explosions
remains elusive. SNe Ia explosion models are frustrated by two main
issues, the uncertainties about thermonuclear flame physics and the 
progenitor evolution. 

Currently,  three explosion scenarios dominate the theoretical models
of SNe Ia. (1) The first (referred to as the ``single-degenerate model")
consists of a  C-O WD near the Chandrasekhar mass
accreting hydrogen or helium from a non-degenerate companion until
it reaches a mass at which the core carbon ignites.  If the subsequent burning
front accelerates to become a detonation in the outer layers of 
the WD, a ``delayed detonation" results. If the burning front remains 
subsonic, the result is a ``deflagration." These single-degenerate models
account for observed inhomogeneities through variations in the propagation of 
the burning front due to density and/or compositional differences in 
the C-O WD progenitor. (2) The second scenario
(referred to as the ``double-degenerate model")
merges two C-O WDs, with the more 
massive WD accreting the companion.
The range of masses of these explosions has been suggested to vary from 
1.2 to 1.8 $M_\odot$, explaining  the observed inhomogeneities of 
SNe Ia. (3) The third scenario (referred to as the ``sub-Chandrasekhar model")
consists of a lower-mass C-O WD accreting a 
helium shell, which becomes thick enough to produce a helium shell 
detonation. Observed differences among 
SNe Ia are explained in this model by the different nucleosynthesis
that results when the progenitor mass varies from 0.65 to 1.1 $M_\odot$.

The pros and cons of these scenarios are not the subject of this paper; 
for reviews see Branch et al. (1995) and Livio (2000). It is worth noting,
however, that Branch (2001) concludes that recent developments favor the 
single-degenerate model for SNe Ia. Years of tuning the single-degenerate
models to achieve a match between observed and computed spectra and light
curves have also resulted in a list of distinct properties which the ``successful"
SN Ia explosion model must possesses. Successful ones include 
deflagration model W7 (Nomoto, Thielemann, \& Yokoi 1984) 
and delayed-detonation model DD4 (Woosley \& Weaver 1994; 
Pinto \& Eastman 2001, hereafter
PE2001). 

PE2001 used the DD4 model and its variants to study the correlation
between luminosity and light-curve width among SNe Ia [similar studies
have been conducted by H\"oflich, Khokhlov, \& Wheeler (1995) and
H\"oflich \& Khokhlov (1996)].
Their DD4 models yield $^{56}$Ni ranging from 0.27 to 0.90 $M_\odot$,
and peak absolute $V$ from $-18.90$ to $-19.78$ mag. 
They are able to reproduce the observed absolute magnitude 
vs. $\Delta m_{15}(B)$ relation [although in their Figure
2 they used $\Delta m_{15}(B)$ measured from the modeled $V$-band
light curve, i.e., it is actually $\Delta m_{15}(V)$]
and found it to be a natural consequence of the radiation transport
in SNe Ia, with the $^{56}$Ni production and the $\gamma$-ray 
escape fraction during the explosion being the dominant factors. 
For example, more $^{56}$Ni leads to more heating, higher temperatures, 
less efficient cooling, and hence broader and brighter light curves, 
while larger $\gamma$-ray escape fractions (e.g., models with higher
overall velocities and/or with $^{56}$Ni distributions extending
to higher velocities) lead to narrower
light curves. Since larger $^{56}$Ni production and higher velocities 
tend to produce opposite effects on light curves of SNe Ia, the solution
is not unique for a particular SN unless additional constraints 
are obtained from observations.

We are particularly interested in PE2001's model DD3 (initially studied
by Woosley \& Weaver 1991), a variant of model DD4/90 which 
has a $^{56}$Ni production of 0.90 $M_\odot$, a peak $V$ absolute
magnitude of $-19.78$, and a $\Delta m_{15}(V)$ of 0.87 mag.
Compared to model DD4/90, model DD3 has greater production of
$^{56}$Ni (0.96 $M_\odot$), a larger total amount of burning,
and higher velocities for the ejecta; it burns 10\% more mass 
to the Si-group or above and has 10\% greater kinetic energy than
model DD4/90. In spite of its high $^{56}$Ni mass, the light curve
from model DD3 is considerably narrower than that of model DD4/90.
The higher kinetic energy in the explosion of model DD3 results
in a lower column depth, allowing $\gamma$-rays to escape 
more easily. At around maximum optical brightness, $\sim$6\% of the decay energy in model
DD4/90 escapes directly as $\gamma$-rays, increasing to 35\% by
$t \approx$ 20 days. In model DD3, the escape fraction is more than
double that of model DD4/90 around maximum, at $\sim$14\%,
rising to 46\% by $t \approx $ 20 days. A shortened time to $\gamma$-ray
transparency for model DD3 leads to a faster rising phase,
a narrower peak, and a
greater decline rate at late times. The increased
$\gamma$-ray escape is,
however, more than compensated by the faster rise and narrower
peak, and the luminosity at peak ($M_V = -19.90$ mag) is even brighter
than the difference one might predict based on consideration of the
difference in total $^{56}$Ni mass. Model DD3 has the same postmaximum
decline rate as model DD4/90 [$\Delta m_{15}(V)$ = 0.87 mag].

We propose that the comparison of SN 2000cx to SN 1991T is analogous
to that of model DD3 to model DD4/90.  SN 2000cx and SN 1991T have
very similar postmaximum decline rates $\Delta m_{15}(B)$.
Since SN 2000cx has much higher expansion velocities for both the 
intermediate-mass and iron-peak elements than SN 1991T does (Figure
18), it has a smaller column density and 
a larger $\gamma$-ray escape fraction. The similar 
$\Delta m_{15}(B)$ for the two SNe thus indicates that SN 2000cx
synthesized more $^{56}$Ni during the explosion than 
did SN 1991T. Mazzali, Danziger, \& Turatto (1995) estimated that 
$\sim 1M_\odot$ of $^{56}$Ni was produced in the explosion of 
SN 1991T, so SN 2000cx may have synthesized more than 1$M_\odot$ of
$^{56}$Ni, the largest ever found for a SN Ia. In spite of 
its higher $^{56}$Ni mass, the light curve is considerably
narrower than that of SN 1991T because the higher kinetic energy 
in SN 2000cx results in a lower column depth, allowing 
$\gamma$-rays to escape more easily. The faster rising phase,
and the greater late-time decline rate in the $BVRI$ passbands 
(\S 2),  are all indications of a larger $\gamma$-ray
escape fraction.

Spectroscopically, the strong resemblance of the premaximum 
spectra of SN 2000cx
to those of SN 1991T provides additional evidence that the two 
SNe are closely related. The prominence of Fe III lines and the
lack of strong Si II, S II, and Ca II lines
at these early times is explained mainly by a composition effect (Filippenko
et al. 1992b; Ruiz-Lapuente et al. 1992; Jeffery et al. 1992), a 
temperature effect (Nugent et al. 1995), or both (Mazzali, Danziger,
\& Turatto 1995). The spectra of SN 2000cx indicate that both the iron-peak
and the intermediate-mass elements are
moving at high expansion velocities in the outer layers of the ejecta (Figure 18),
suggesting that both the composition and the temperature
significantly affect the formation of the spectral lines at this
time. In other words, the strong Fe III lines may be caused by the 
composition abnormality of iron-peak elements in the outer layers
of the ejecta, but the lack of strong lines of the intermediate-mass elements 
is caused by the high radiation temperature, not the physical absence
of such elements.

As discussed in \S 3, although SNe 2000cx and 1991T
have similar premaximum
spectra, their overall spectral evolution is quite different. This is almost
certainly caused by differences in the 
$^{56}$Ni mass and the expansion velocities of the ejecta. 
SN 2000cx has a higher $^{56}$Ni mass and  hotter
ejecta, but the faster expansion velocities 
yield a larger $\gamma$-ray escape fraction so the continua
of the observed spectra are not particularly blue (in fact, the 
color at the earliest epochs is redder than that of SN 1991T; 
\S 3). 
The temperature of the photosphere of SN 2000cx must have been dropping
from $t$ = $-$3 days to a week after maximum brightness, as the features of 
the intermediate-mass elements (in particular Si and S) grow stronger.

The peculiar photometric and spectroscopic evolution of SN 2000cx
during $t$ = 6 to 15 days [i.e., the transition phase in the $B$-band light
curve, the plateau phase in the $(B - V)_0$ color, the bluest part 
of the $(V - R)_0$ and $(V - I)_0$ color, the prominent Fe III and 
Si II lines, and the blue continuum], might be due to  changes 
in the ratio of  the $\gamma$-ray escape fraction of SN 2000cx 
relative to SN 1991T.
As demonstrated by models DD3 and DD4/90, the
ratio of the $\gamma$-ray escape fraction of model DD3 to that of 
model DD4/90 is $\sim230$\% around maximum, and it drops to only 
$\sim130$\% at $t \approx$ 20 days. For SN 2000cx, the drop of this 
ratio means its continuum formation does not differ much from that of
other SNe Ia, and its high intrinsic radiation temperature
(which is supported by the strength of the Fe III lines) 
yields a much bluer continuum. This is also the reason why the $(V - R)_0$
and $(V - I)_0$ colors are so blue;  they are sensitive
to the continuum shape. The $(B - V)_0$
color stays nearly constant during this period because, 
as discussed in \S 3,
the change in the continuum shape is compensated by variations 
in the spectral features. The decrease of the ratio of the $\gamma$-ray
escape fraction may also be
the reason for the transition phase in the $B$-band light curve.

The late-time spectral and photometric evolution of SN 2000cx can
also be understood in terms of a higher radiation temperature and
a larger $\gamma$-ray escape fraction. The higher radiation temperature
explains the unusual blue color evolution at late times, while 
the larger $\gamma$-ray escape fraction is responsible for the 
steeper decline rate in the late-time $BVRI$ light curves. 

We should note that even though the delayed detonation model
DD3 (a single-degenerate model) seems to account for the
observations of SN 2000cx quite well,
the essence of the model (high $^{56}$Ni mass and high 
expansion velocity of the ejecta) could be produced in other
scenarios as well. For example, Filippenko et al. (1992b) suggested a 
double-detonation model for SN 1991T,  in which a mildly
sub-Chandrasekhar mass WD is nearly completely incinerated by 
detonation waves propagating inward and outward from the base of 
an accumulated helium layer.
This model is capable of producing a large
amount of $^{56}$Ni because of the complete burning to iron-peak
elements by detonation. The outward detonation also produces material
moving at high expansion velocities. However, the presence of 
intermediate-mass elements at high velocities, and the persistent
strong Si II lines during $t$ = 7 to 20 days,  seem to contradict
nucleosynthesis from complete detonations.  Fisher et al. (1999)
used a super-Chandrasekhar explosion from the merger of
two WDs as the origin of SN 1991T.
The expected strong explosion from a super-Chandrasekhar WD could
produce large amounts of $^{56}$Ni, and a high-velocity iron-peak core
surrounded by a small mass of intermediate-mass elements; this 
encounters a surrounding low-density mass of carbon and oxygen
of the companion WD which decelerates the intermediate-mass
elements and forces them into a narrow velocity interval. Our spectra
of SN 2000cx indicate that the intermediate-mass elements are indeed
confined to a fairly restricted layer characterized by velocities
around 12,000 km s$^{-1}$.  Although the necessity of  a 
super-Chandrasekhar explosion for SN 1991T is questionable 
after the Cepheid distance to its host galaxy NGC 4527 was
measured (Saha et al.  2001) and SN 1991T was found to be only mildly
overluminous compared to normal SNe Ia, the super-Chandrasekhar explosion
may be a
viable model for SN 2000cx. There are, however, few studies of
this model in the literature, so a detailed comparison
with our observations cannot be made. 

We also emphasize that since there are still relatively large uncertainties
in the theoretical models for SNe Ia (e.g., details about thermonuclear
flame physics and the progenitor evolution),  our comparison of SN 2000cx
and SN 1991T,  and the analog of the comparison of model DD3 and 
DD4/90, are only qualitative. A comprehensive quantitative study
should involve detailed modeling of spectra and multicolor light curves
of model DD3,  and comparison with the observations of SN 2000cx. This
is beyond the scope of the current paper. It is worth noting, however,
that the comparison between model DD3 and DD4/90 is done in the $V$ band
in PE2001, while our comparison concentrated mostly on the $B$-band
observations of SN 2000cx. It is unclear whether the correlations 
in the $V$ band reported by PE2001 apply to the $B$ band as well.
Moreover,  SN 2000cx is very distinct from SN 1991T 
in the $V$, $R$, and $I$ passbands (Figures 4, 5, and 6), so 
they may be quite different objects.

\subsection{Implications of the SN 2000cx Observations}

If we accept the theoretical connection between 
SN 2000cx and SN 1991T discussed above, 
SN 2000cx should have absolute magnitudes brighter than those 
of SN 1991T  by about 0.1 mag as indicated by the difference
in models DD3 and DD4/90. Adopting the absolute magnitudes of 
SN 1991T from 
Saha et al. (2001) [they assumed a reddening of $E(B - V)$ =
0.2$\pm$0.1 mag to SN 1991T], the absolute magnitudes for SN 2000cx are 
$M_B = -19.96\pm0.36$, $M_V = -19.95\pm0.29$ mag, and $M_I = -19.55\pm0.21$
mag.

The absolute magnitudes derived from distance estimates
and the light-curve fitting techniques (Table 6) are systematically
fainter than the theoretical expectations. 
This indicates that NGC 524 may be more distant than 
estimated from its
recession velocity. On the other hand,  given the bad fits
discussed earlier,
the difference between the estimates from the 
theoretical models and those from the light-curve fitting 
techniques is not surprising, and the estimates are still
consistent within their uncertainties.
The closest match to the estimates 
from the theoretical models are those derived from the recession-velocity 
distance with
$H_0 = 58.5$ km s$^{-1}$ Mpc$^{-1}$, from the stretch method fit to
the postmaximum data points, and from the luminosity
vs. $\Delta m_{15}(B)$ relation. 

If the expansion velocity of a SN Ia affects its light-curve shape, 
as suggested by PE2001, the current luminosity vs. light-curve width
relation should be revised by taking into account the expansion
velocities of SNe Ia. Since higher $V_{exp}$ reduces the light-curve
width and increases the $\Delta m_{15}(B)$ value\footnote{Although the $\Delta m_{15}$
value of model DD3 is the same as that of model DD4/90, it is relatively
too large for the higher luminosity and larger $^{56}$Ni mass of model DD3.}, 
given the same $\Delta m_{15}(B)$ value the corrected luminosity should 
be higher for a SN Ia with higher $V_{exp}$.  There is a trend that 
SNe Ia in late-type galaxies have higher $V_{exp}$ (Filippenko 1989; Branch \&
van den Bergh 1993) and higher luminosity (Hamuy et al. 2000)
than those in early-type galaxies, so the $V_{exp}-$corrected 
luminosity vs. light-curve width relation will have a steeper slope than
currently assumed. 

We emphasize again that the above conclusion is based on the 
result by PE2001 that the expansion velocity affects the light-curve formation. 
More theoretical studies  need to  be conducted to verify this 
result, and to provide a quantitative guide on how to correct the current
luminosity vs. light-curve width relation.

One interesting question is  how frequent SN 2000cx-like
objects are. SN 2000cx is clearly a peculiar SN Ia, perhaps
closely related to SN 1991T-like objects. Li et al. (2001) 
studied a well-understood sample of SNe Ia and concluded that 
the rate of peculiar SNe Ia such as SN 1991T and SN 1991bg (Filippenko
et al. 1992a) may
constitute a significant fraction of the observed SNe Ia ($\sim$ 20\%
are SN 1991T-like and $\sim$ 16\% are SN 1991bg-like). We failed
to find a SN Ia similar to SN 2000cx, however,  both in the 
literature and in our photometric database, so objects like 
SN 2000cx may be very rare and originate only under 
extreme conditions. 
Such mavericks, however, may reveal trends that are otherwise buried
in observational uncertainties. 

Another interesting question is whether the peculiarity of SN 2000cx
is related to the nature of its environment. SN 2000cx is located 
very far away from the nucleus of its host galaxy, where the 
metallicity is likely to be low given the abundance gradients observed
in galaxies (Henry \& Worthey 1999). Hamuy et al. (2000) 
suggested that metal-poorer environments produce the most luminous
SNe Ia, which is supported by SN 2000cx if it is indeed more luminous
than SN 1991T. On the other hand, Ivanov, Hamuy, \& Pinto (2000) studied
the radial distribution of SNe Ia and showed that there is no indication
of systematic changes of the luminosity of SNe Ia with radial distance,
suggesting that the metallicity effects on SNe Ia are likely to be small.
 
\section{CONCLUSIONS}

(1) SN 2000cx is a very peculiar object -- indeed, unique among all known
SNe Ia. The light curves cannot be fit well by any of the fitting
techniques currently available (e.g., MLCS and the stretch method).
There is an apparent asymmetry in the rising and declining
parts of the $B$-band light curve, while there is a unique ``shoulder''-like
evolution in the $V$-band light curve. The $R$-band and $I$-band light curves
have relatively weak second maxima.  In all $BVRI$ passbands the 
late-time decline rates are relatively large compared to other SNe Ia.

(2) SN 2000cx has the reddest $(B - V)_0$ color before $t \approx$ 7 days among
several SNe Ia, and it subsequently has a peculiar plateau phase where $(B - V)_0$
remains at 0.3 mag until $t$ = 15 days. The late-time $(B - V)_0$ evolution
of SN 2000cx is found to be rather blue, and is inconsistent with the fit
proposed by Lira (1995) and Phillips et al. (1999). SN 2000cx also has very 
blue $(V - R)_0$ and $(V - I)_0$ colors compared to other SNe Ia.

(3) Our earliest spectrum of SN 2000cx ($t = -3$ days) reveals remarkable
resemblance to those of SN 1991T-like objects, with prominent Fe III lines and
weak Si II lines. As in the case of SN 1991T, Si~II lines strengthened around
the time of maximum brightness. However, the subsequent spectral evolution of
SN 2000cx is quite different from that of SN 1991T.  The Fe~III and Si~II lines
remain strong, and the Fe~II lines remain weak, in the spectra of SN 2000cx
until $t \approx$ 20 days, indicating that the excitation stages of iron-peak
elements change relatively slowly in SN 2000cx compared to other SNe Ia, and
suggesting that the photosphere of SN 2000cx stays hot for a long time. Both
iron-peak and intermediate-mass elements are found to be moving at very high
velocities in SN 2000cx. The $V_{exp}$ measured from the Si II $\lambda$6355
line shows a peculiar (nearly constant) evolution.

(4) We find that the delayed detonation model DD3 (Woosley \& Weaver 1991) 
investigated by Pinto \& Eastman (2000b) accounts for the observations
of SN 2000cx rather well. This model suggests that SN 2000cx is similar
to SN 1991T, but with a larger $^{56}$Ni production 
and a higher kinetic energy (i.e., greater expansion velocity for the 
ejecta). We emphasize that because of uncertainties in the current theoretical 
models for SNe Ia, various views should be considered. For example, 
the big difference between SN 2000cx and SN 1991T in their $V, R, $ and $I$ 
light curves may suggest that they are two very different objects.

\acknowledgments

We thank the staffs of the Lick Observatory and Wise Observatory for 
their assistance, and we acknowledge useful conversations with K. Krisciunas
and M. Davis. The work of A.V.F.'s group at U. C. Berkeley is
supported by the National
Science Foundation grant AST-9987438, as well as by the Sylvia and Jim
Katzman Foundation. KAIT was made possible by
generous donations from Sun Microsystems, Inc.,
the Hewlett-Packard Company, AutoScope Corporation, Lick Observatory, 
the National Science Foundation, the University of California,
and the Katzman Foundation.  A.V.F. is grateful to the Guggenheim 
Foundation for a Fellowship. 
Astronomy at the Wise Observatory is supported by grants from the Israel
Science Foundation. The Wise Observatory queue-observing program is made 
possible by the devoted help of the Wise Observatory staff.

\newpage

\begin{figure}
\caption{\emph{V}-band KAIT image of the field of SN 2000cx in NGC 524, taken on 2000
July 25. The field of view is
6$\farcm$7 $\times$ 6$\farcm$7. The four local standard stars are marked (1 -- 4).}
\label{1}
\end{figure}

\begin{figure}
\caption{The \emph{B, V, R,} and \emph{I} light curves of SN 2000cx. The open
circles are the KAIT measurements and the solid circles are the WO
data. For most of the points the uncertainties are smaller than the
plotted symbols.}
\label{2}
\end{figure}

\begin{figure}
\caption{The \emph{B}-band light curve of SN 2000cx, together with those of
SN 1991T (Lira et al. 1998), SN 1991bg (Leibundgut et al. 1993),
SN 1992bc (Hamuy et al. 1996b), SN 1992al (Hamuy
et al. 1996b), and SN 1994D (Richmond et al. 1995). All light curves are shifted
in time and peak magnitude to match that of SN 2000cx. The inset shows the
comparison near maximum. }
\label{3}
\end{figure}

\begin{figure}
\caption{Same as Figure 3 but for the \emph{V} light curve.}
\label{4}
\end{figure}

\begin{figure}
\caption{Same as Figure 3 but for the \emph{R} light curve. The \emph{R}
light curve of SN 1991bg comes from Filippenko et al. (1992a). }
\label{5}
\end{figure}

\begin{figure}
\caption{Same as Figure 3 but for the \emph{I} light curve. The \emph{I}
light curve of SN 1991bg comes from Filippenko et al. (1992a). }
\label{6}
\end{figure}

\begin{figure}
\caption{The $(B - V)_0$ color curve of SN 2000cx, together with those
of SN 1991T [dereddened by $E(B - V)$ = 0.13 mag],
SN 1992bc [dereddened by $E(B - V)$ = 0.02 mag], SN 1992al [dereddened
by $E(B - V) = 0.04$ mag], and SN 1994D [dereddened by $E(B - V) = 0.04$
mag]. The wide line labeled ``L-P law" is the fit by Lira (1995) and
Phillips et al. (1999).}
\label{7}
\end{figure} 

\begin{figure}
\caption{Same as in Figure 7 but for the $(V - R)_0$ color curve.}
\label{8}
\end{figure}

\begin{figure}
\caption{Same as in Figure 7 but for the $(V - I)_0$ color curve.}
\label{9}
\end{figure}

\begin{figure}
\caption{Montage of spectra of SN 2000cx obtained at $t <$ 30 days.
The phases marked are relative to
the date of $B$ maximum (those denoted with a ``p" were obtained in
spectropolarimetric mode). To improve clarity, the spectra have been
shifted vertically by arbitrary amounts. The spectra have been corrected
for reddening [$E(B - V) = 0.08$ mag, see text for details] and redshift
(2421 km s$^{-1}$).}
\label{10}
\end{figure}

\begin{figure}
\caption{Same as in Figure 10 but for spectra obtained at $t > $ 30 days.
The region redward of 7500 \AA\, has been boxcar
smoothed (smoothing box = 5 pixels) for the spectra at days 98, 126,
and 148.
}
\label{11}
\end{figure}

\begin{figure}
\caption{The spectrum of SN 2000cx at $t$ = $-3$ days, shown with
comparable-phase spectra of SN 1994D and SN 1997br. All the spectra
illustrated here and in subsequent figures have been dereddened and deredshifted.
See text for sources of line identifications.}
\label{12}
\end{figure}

\begin{figure}
\caption{Same as in Figure 12 but for spectra near maximum brightness.}
\label{13}
\end{figure}

\begin{figure}
\caption{Same as in Figure 12 but for spectra around $t$ = 14 days.
The features marked with a ``$\oplus$" in the spectrum of SN 1997br
are caused by telluric absorption bands. The four dotted vertical
lines are (from  left to right) the expected positions
of the Fe III $\lambda$4404, Fe II $\lambda$4555,
Fe III $\lambda$5129, and Fe II $\lambda$5169 absorptions. Note the strong
Fe III and weak Fe II lines in the spectrum of SN 2000cx.}
\label{14}
\end{figure}

\begin{figure}
\caption{Same as in Figure 12 but for spectra around $t$ = 20 days.
The two dotted vertical lines are the expected positions of  the Fe II
$\lambda$4555 and Fe II $\lambda$5169 absorptions. Weak Fe II lines
begin to develop in the spectrum of SN 2000cx.}
\label{15}
\end{figure}

\begin{figure}
\caption{Same as in Figure 12 but for spectra around $t$ = 30 days.
The features marked with a ``$\oplus$" in the spectrum of SN 1997br
are caused by telluric absorption bands. The two dotted vertical lines
are the expected positions of the Fe II $\lambda$4555 and Fe II $\lambda$5169
absorptions. The Fe II lines are apparent in the spectrum of SN 2000cx.}
\label{16}
\end{figure}
\clearpage

\begin{figure}
\caption{Same as in Figure 12 but for spectra at late times.}
\label{17}
\end{figure}

\begin{figure}
\caption{Evolution of the expansion velocity as deduced from the
minima of the Fe III $\lambda$4404, Fe III $\lambda$5129, Fe II $\lambda$4555,
S II $\lambda$5468, S II $\lambda\lambda$5612, 5654, and Si II $\lambda$6355
absorption lines for SN 2000cx (solid circles; measured from the spectra
displayed in Figure 10), SN 1991T [open circles; measured from the spectra
in Filippenko et al. (1992b)], and other SNe Ia.
The $V_{exp}$ derived from the Fe II $\lambda$4555, S II $\lambda$5468,
and S II $\lambda\lambda$5612, 5654 lines for SN 1994D are measured
from the spectral series in Filippenko (1997). The expansion velocities
measured from the Si II $\lambda$6355 line are adopted from Branch et al.
(1983) for SN 1981B, Phillips et al. (1987) for SN 1986G, Wells et al.
(1994) for SN 1989B, Leibundgut et al. (1993) for SN 1991bg, and
Patat et al. (1996) for SN 1994D.
}
\label{18}
\end{figure}

\begin{figure}
\caption{The MLCS fit (left panel) and the stretch method fit
(right panel) for SN 2000cx. The MLCS fit is the worst we have
ever seen. For the stretch method fit, the solid line is the fit to
all the data points from $t$ = $-$8 to 32 days, the dash-dotted line
uses only the premaximum datapoints,  and the dashed line for the postmaximum
datapoints. The three fits give very different stretch factors.}
\label{19}
\end{figure} 

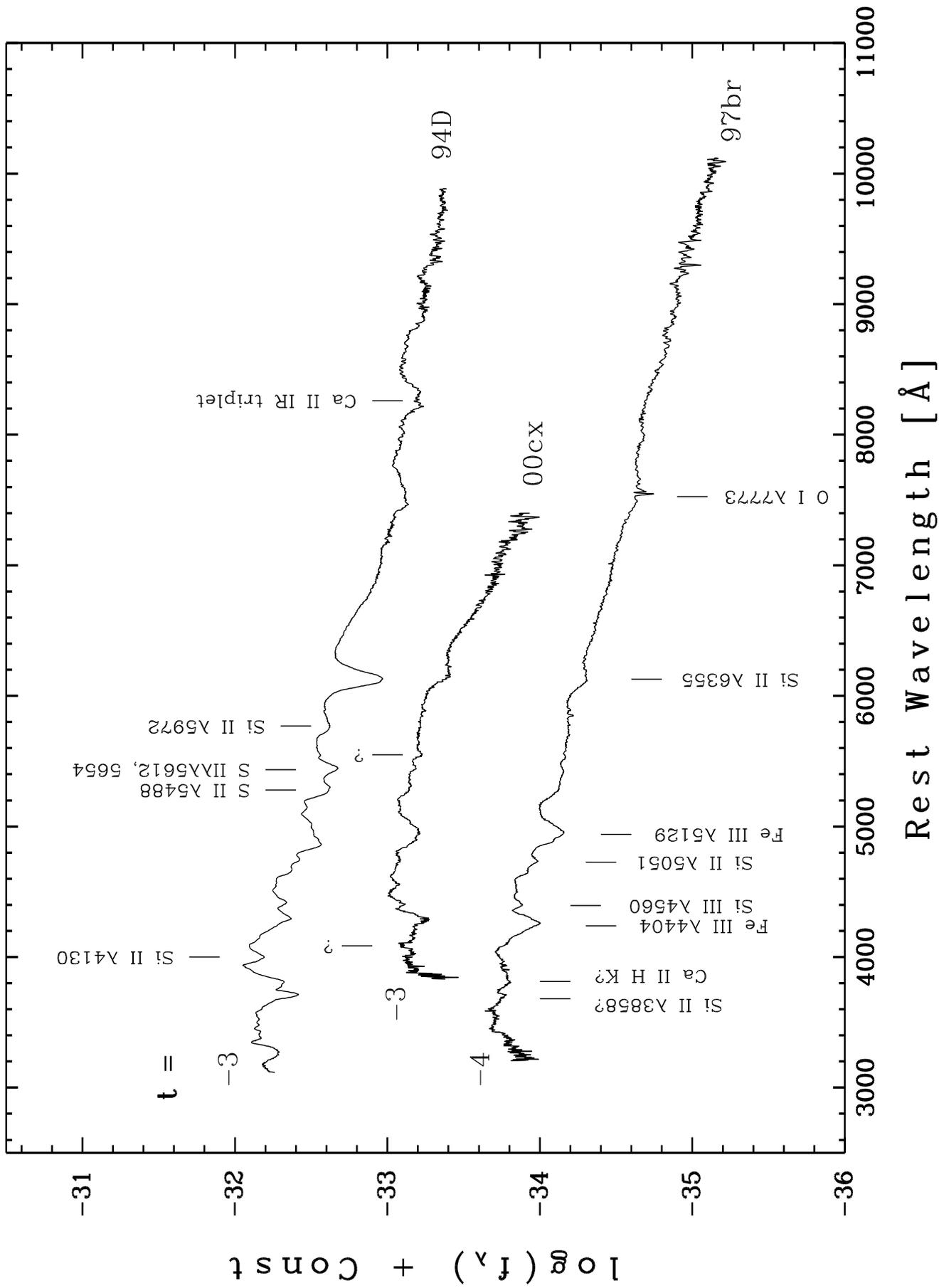
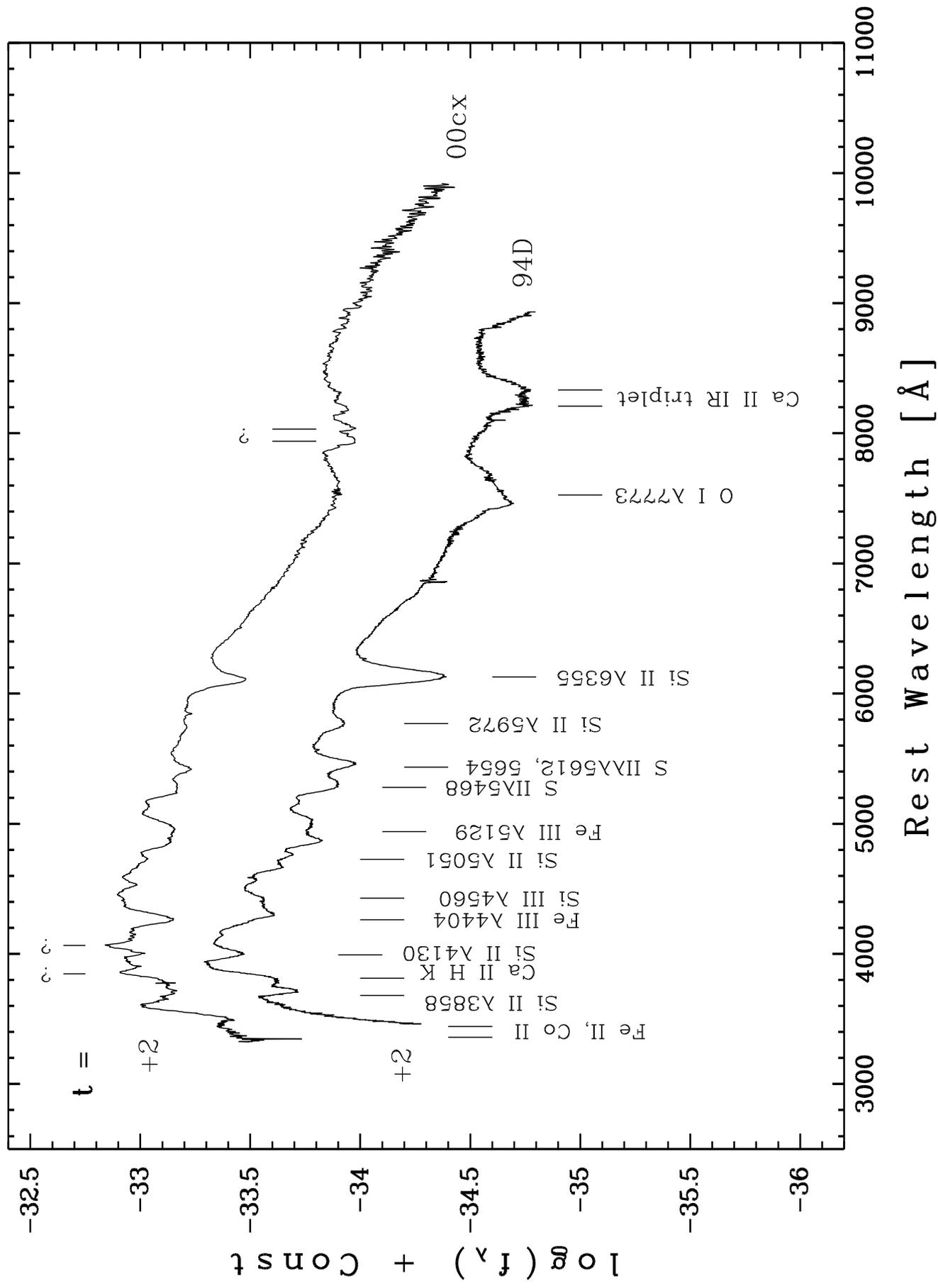
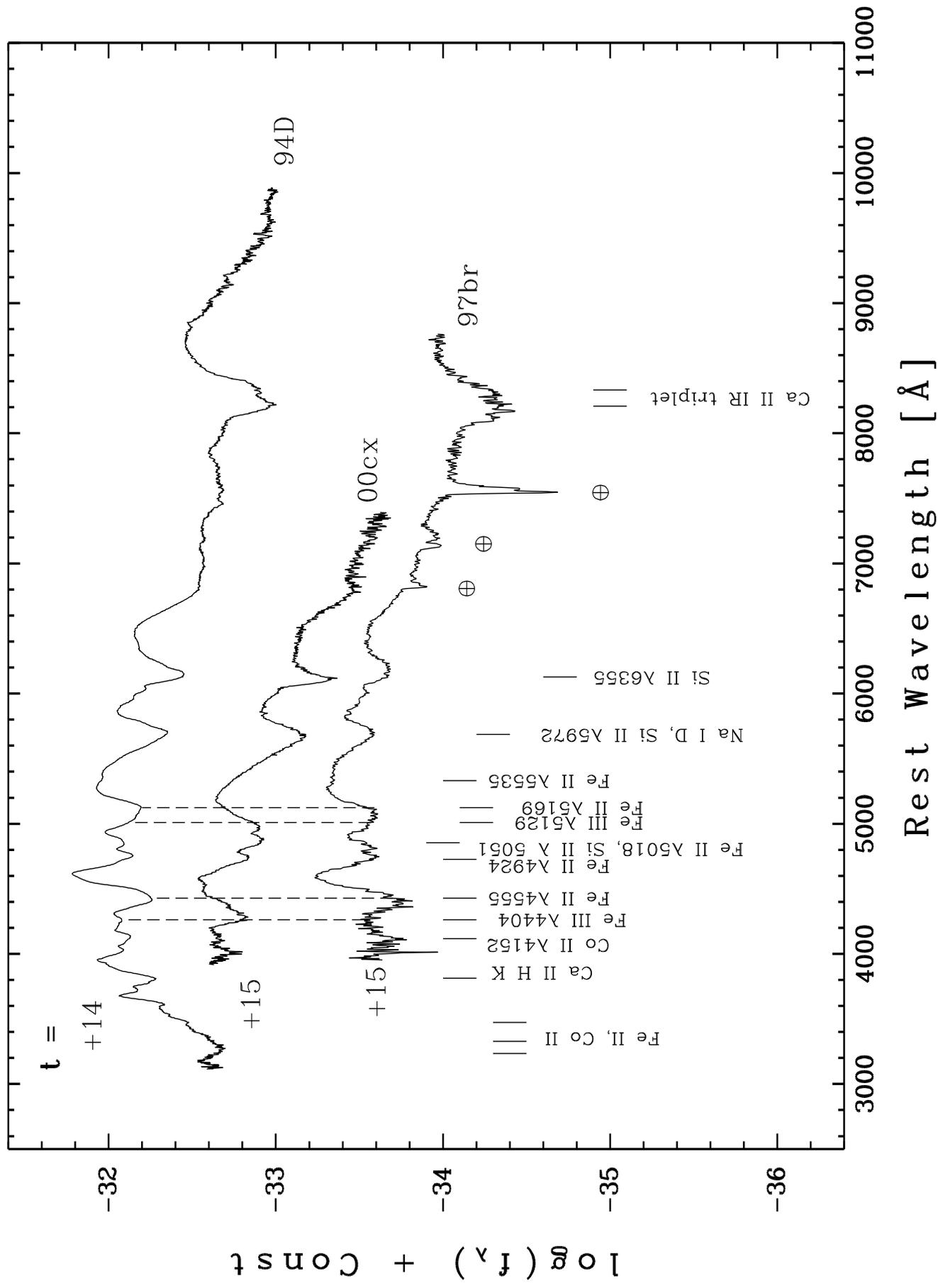
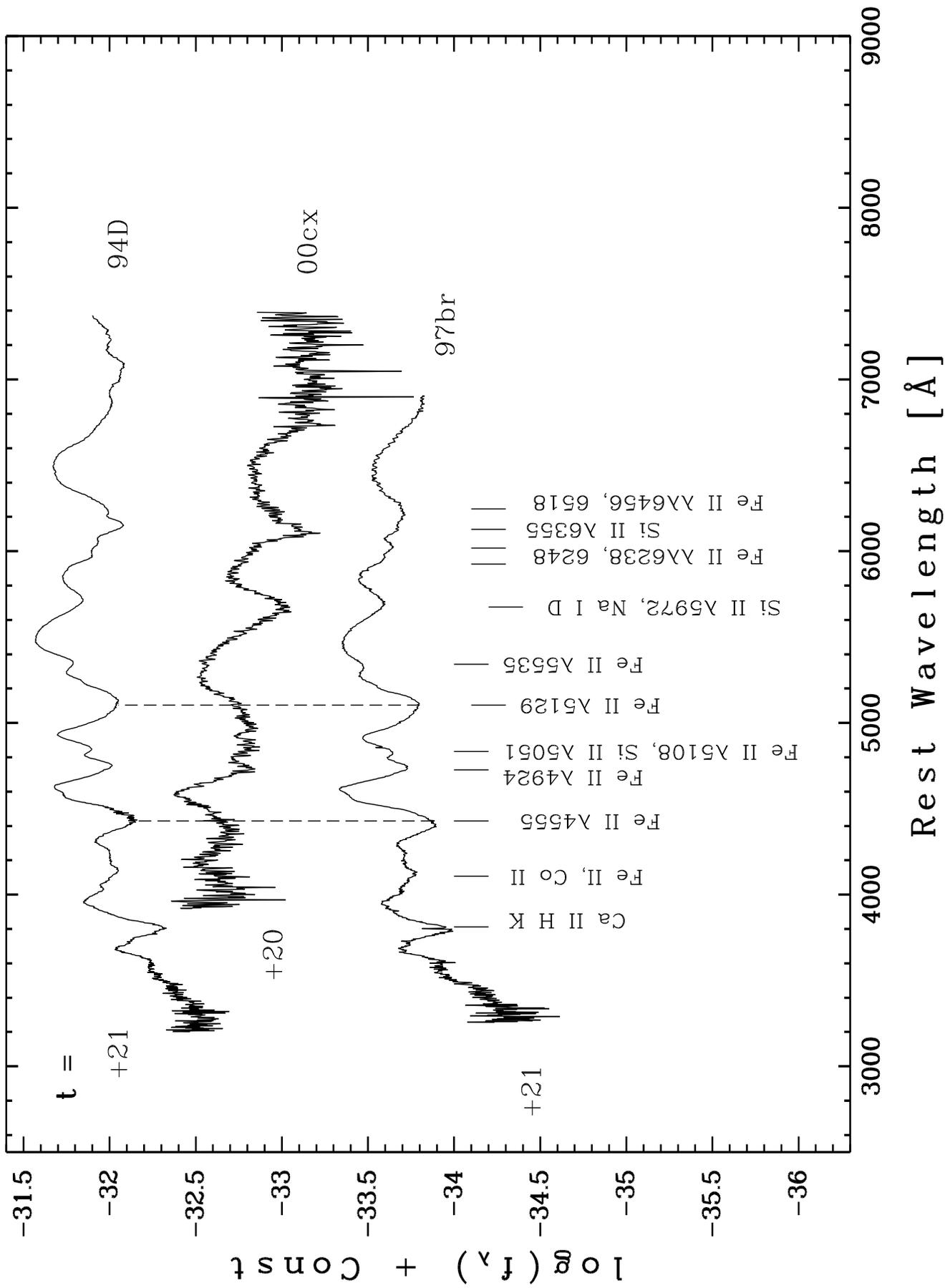
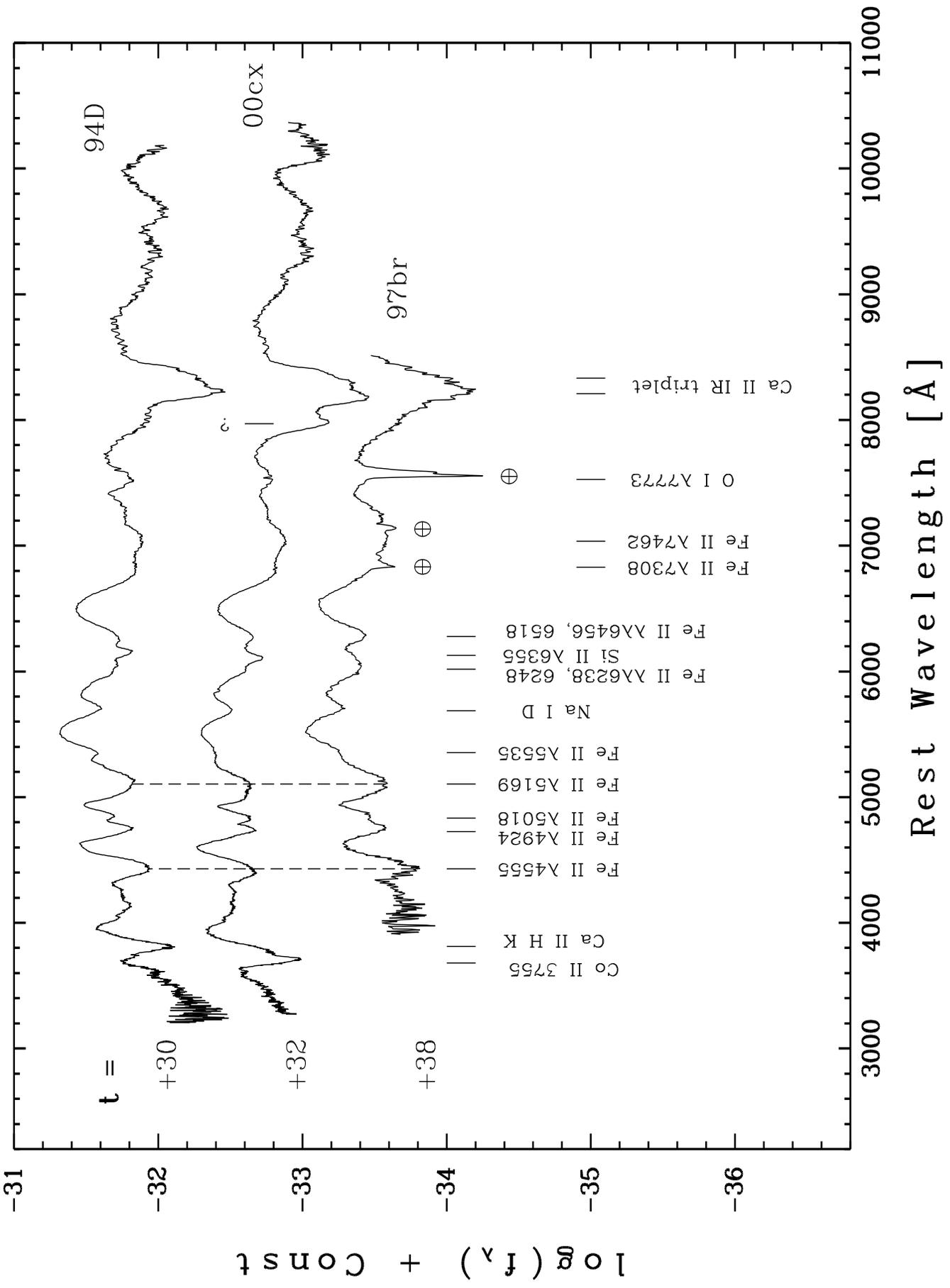
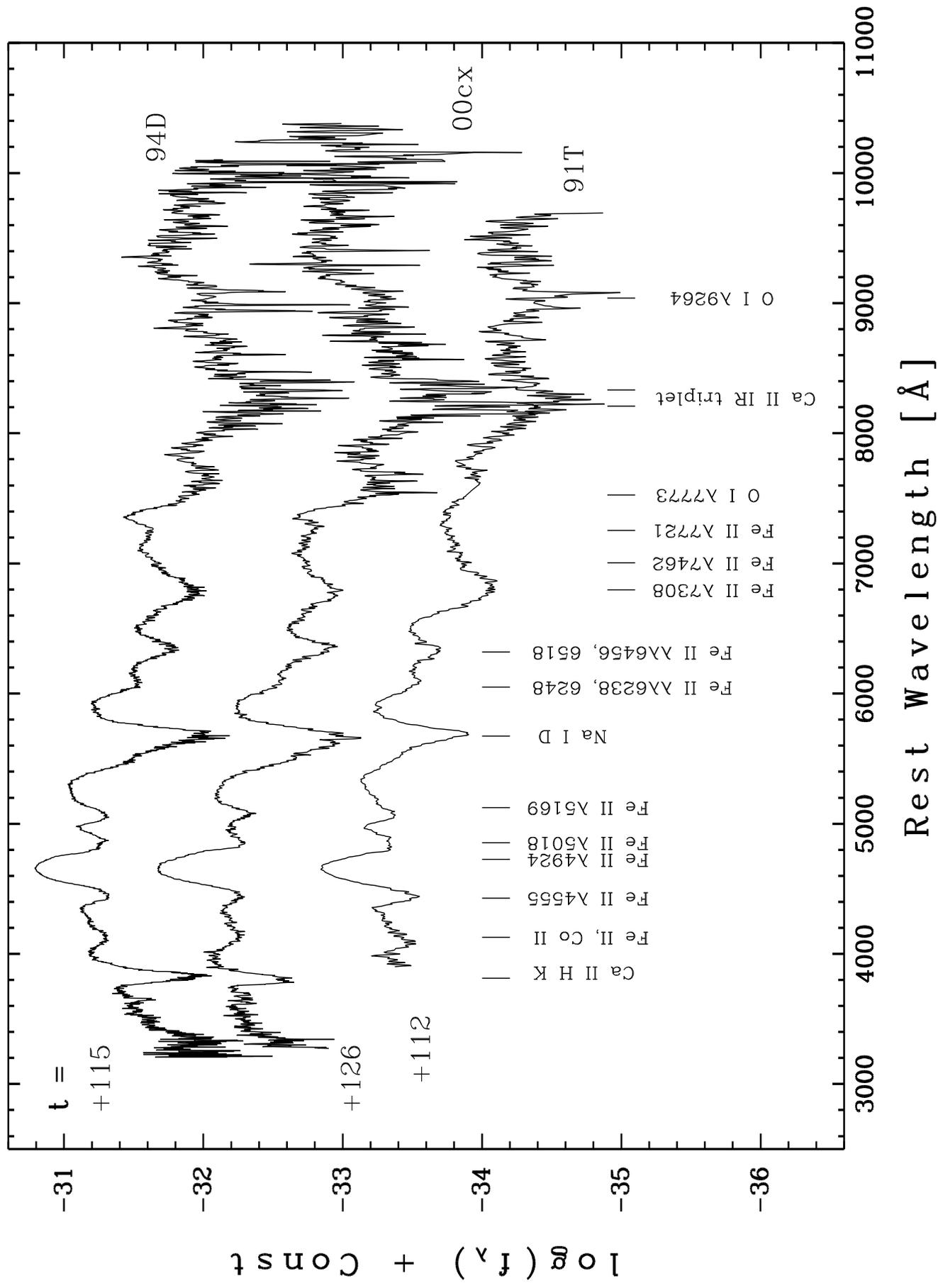
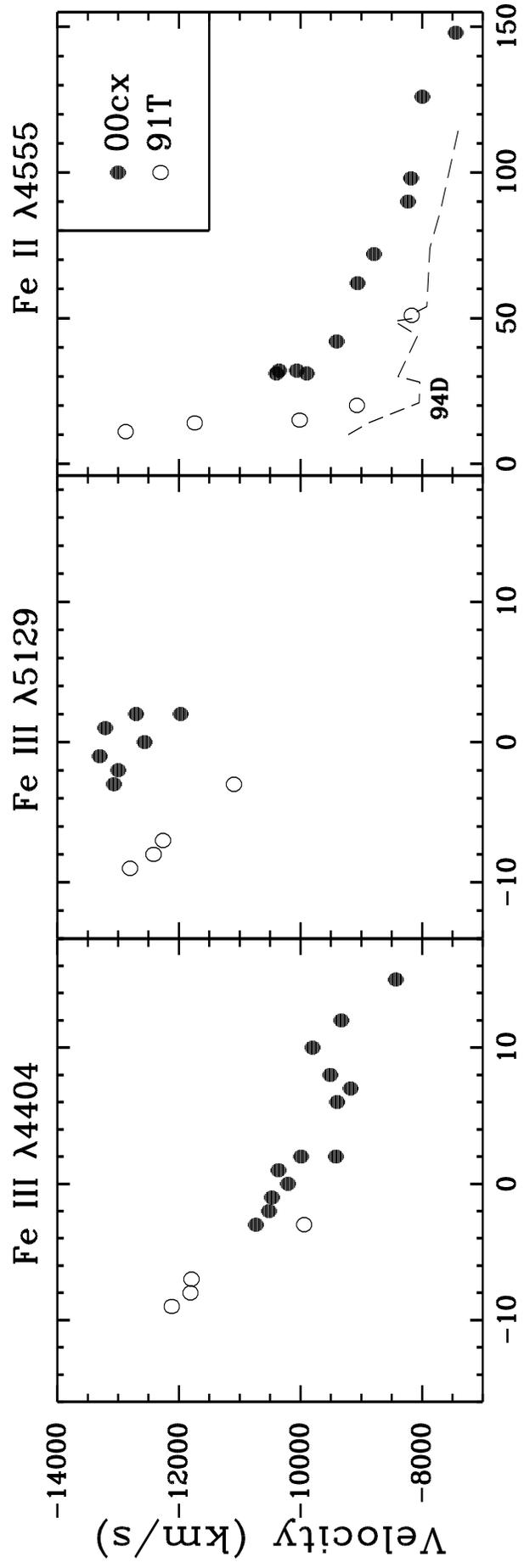
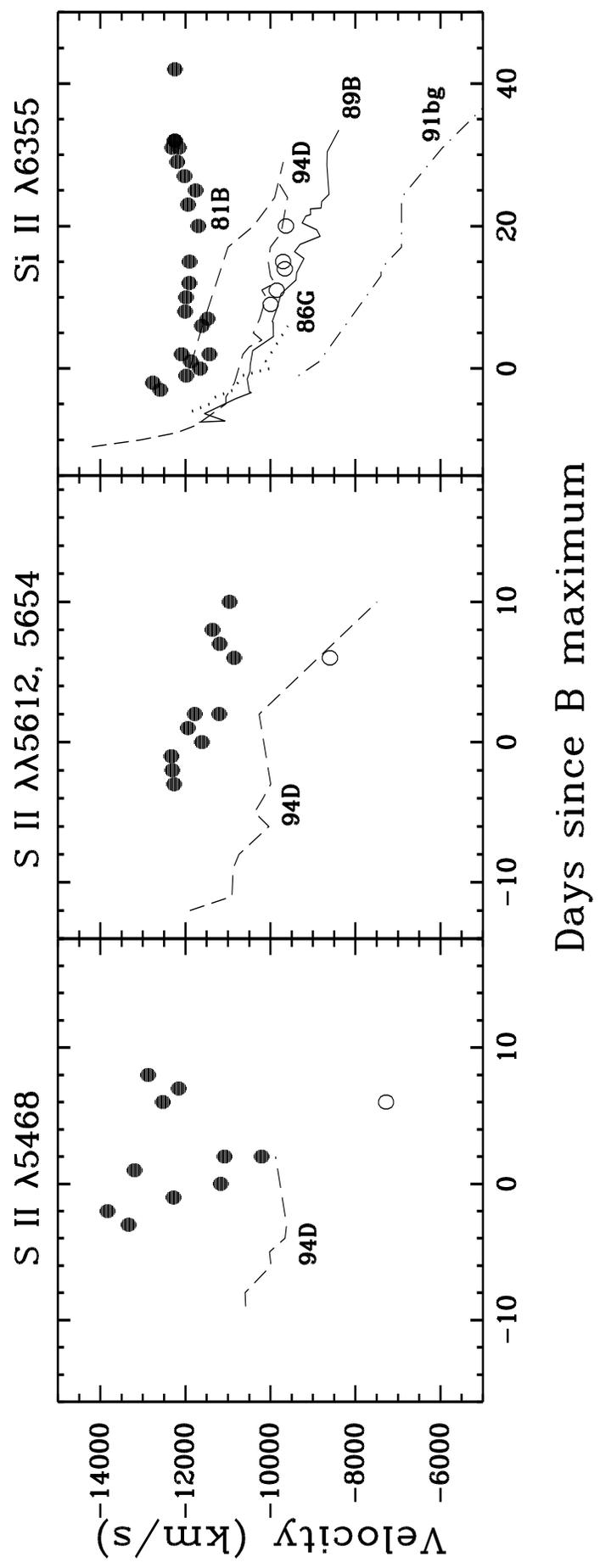
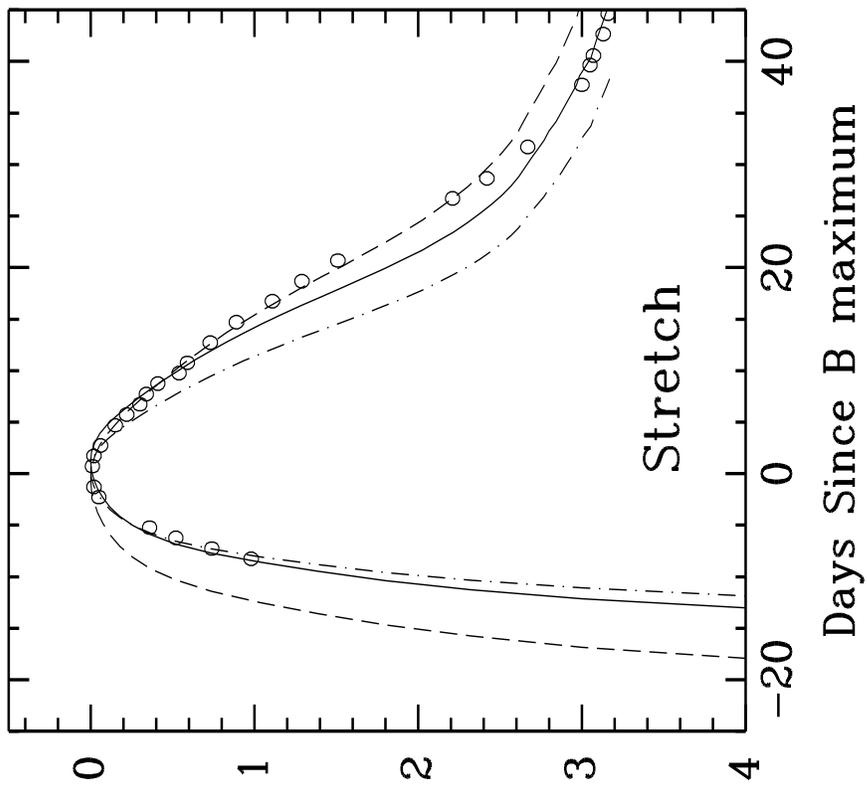
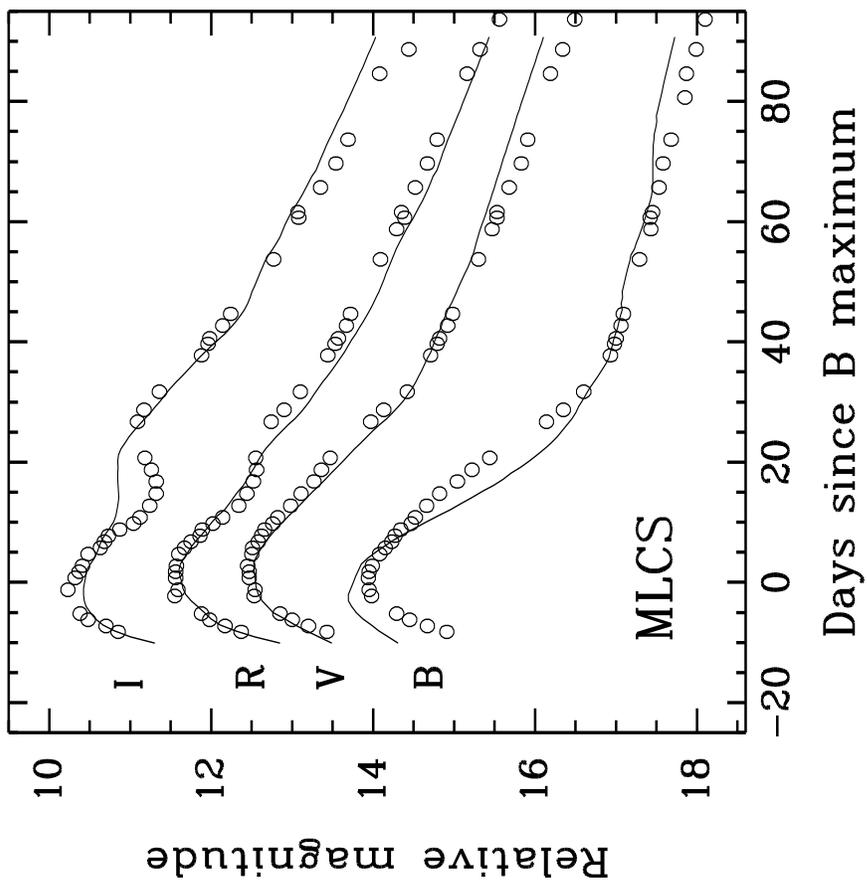
\end{document}